\documentclass[12pt,preprint]{aastex}

\usepackage{emulateapj5}
\usepackage{epsf}

\newcommand{\tx}[1]{\textrm{#1}}

\newcommand{\kms}{km~$\tx{s}^{-1}$}

\newcommand{\be}{\begin{equation}}
\newcommand{\ee}{\end{equation}}

\newenvironment{inlinefigure}{
\def\@captype{figure}
\noindent\begin{minipage}{0.999\linewidth}\begin{center}}
{\end{center}\end{minipage}\smallskip}

\shorttitle{A Wide Field HST Study of CL0024+16. I}
\shortauthors{T. Treu et al.\ }

\begin{document}

\title{A Wide Field Hubble Space Telescope Study of the Cluster
CL0024+16 at $z=0.4$. I: Morphological Distributions to 5 Mpc Radius}

\author{Tommaso Treu\altaffilmark{1}, Richard
S. Ellis\altaffilmark{1}, Jean-Paul Kneib\altaffilmark{1,2}, Alan
Dressler\altaffilmark{3}, Ian Smail\altaffilmark{4}, Oliver
Czoske\altaffilmark{2,6}, Augustus Oemler\altaffilmark{3}, Priyamvada
Natarajan\altaffilmark{5}}

\altaffiltext{1}{California Institute of Technology, Astronomy,
105-25, Pasadena, CA 91125. tt@astro.caltech.edu; rse@astro.caltech.edu}
\altaffiltext{2}{Observatoire Midi-Pyr\'en\'ees, UMR5572, 14 Av. \'Edouard Belin, 31400 Toulouse, France}
\altaffiltext{3}{Carnegie Observatories, 813 Santa Barbara Street, Pasadena, CA 91101}
\altaffiltext{4}{Department of Physics, University of Durham, South Road, Durham DH1 3LE, UK}
\altaffiltext{5}{Department of Astronomy, Yale University, P.O. Box 208101, New Haven, CT 06250}
\altaffiltext{6}{Current address: Institut f\"ur Astrophysik und Extraterrestrische Forschung
Auf dem H\"ugel 71, 53121 Bonn, Germany}

\begin{abstract}

We describe a new wide field Hubble Space Telescope survey of the
galaxy cluster Cl0024+16 ($z\approx0.4$) consisting of a
sparse-sampled mosaic of 39 Wide Field and Planetary Camera 2 images
which extends to a cluster radius of $\sim5$ Mpc. Together with
extensive ground-based spectroscopy taken from the literature,
augmented with over a hundred newly-determined redshifts, this unique
dataset enables us to examine environmental influences on the
properties of cluster members from the inner core to well beyond the
virial radius ($\sim 1.7$ Mpc). We catalog photometric measures for
22,000 objects to $I\ga25$ and assign morphological types for 2181 to
$I=22.5$, of which 195 are spectroscopically-confirmed cluster
members. We examine both the morphology-radius (T-R) and
morphology-density (T-$\Sigma$) relations and demonstrate
sensitivities adequate for measures from the core to a radius of $\sim
5$ Mpc, spanning over 3 decades in local projected density. The
fraction of early-type galaxies declines steeply from the cluster
center to 1 Mpc radius and more gradually thereafter, asymptoting
towards the field value at the periphery. We discuss our results in
the context of three distinct cluster zones, defined according to
different physical processes that may be effective in transforming
galaxy morphology in each.  By treating infalling galaxies as isolated
test particles, we deduce that the most likely processes responsible
for the mild gradient in the morphological mix outside the virial
radius are {\it harassment} and {\it starvation}. Although more data
are needed to pin down the exact mechanisms, {\it starvation} seems
more promising in that it would naturally explain the stellar and
dynamical homogeneity of cluster E/S0s. However, we find significant
scatter in the local density at any given radius outside $\sim$0.5
Mpc, and that the same T-$\Sigma$ relation holds in subregions of the
cluster, independent of location. In this hitherto unprobed region,
where the potential of the cluster is weak, galaxies apparently retain
their identities as members of infalling sub-groups whose
characteristic morphological properties remain intact. Only upon
arrival in the central regions is the substructure erased, as
indicated by the tight correlation between cluster radius and
$\Sigma$.

\end{abstract}


\section{Introduction}

Rich clusters offer an unique laboratory for studying the effects of
the local environment on their constituent galaxies. Evidence has
accumulated in recent years that environmental processes affect both
the star formation and morphological characteristics of galaxies, the
endpoint being the present-day morphology-density relation (Dressler
1980).

However, the nature and timescale of the relevant evolutionary
processes remains unclear. As clusters represent concentrations of
both dark and baryonic matter, the physical interplay between the
observed properties of a galaxy and the cluster potential is
likely complex. An important advance was the recognition that the
hot intercluster gas observed in X-rays may significantly affect
the star formation rate of recently-arrived infalling galaxies
(Gunn \& Gott 1972), possibly triggering star formation by
compressing their gas clouds or quenching star formation by
stripping their gas reservoir (Dressler \& Gunn 1983; Byrd \&
Valtonen 1990; Evrard 1991; Fujita 1998; Abadi et al.\ 1999)

Much progress has followed the connection of Hubble Space Telescope
(HST) imaging of clusters with deep ground based spectroscopy. The
former is essential in providing galaxy morphologies at epochs where
environmental processes can be witnessed, while the latter is
essential in providing diagnostic information on the timescales of
star formation and its recent truncation. The combination has been
most powerful in understanding the origin of the increase with
redshift in the fraction of blue members seen in cluster cores between
$z=0$ and $z\sim0.4$ (Butcher \& Oemler 1978, 1984). Several HST-based
studies have demonstrated that the blue galaxies seen at $z>$0.3 are
largely star-forming spirals which are noticeably absent in
present-day clusters. It has been proposed these are recent arrivals
in the cluster, perhaps observed immediately prior to a removal of
their gas supply. Importantly, some red cluster galaxies in the same
clusters show evidence of recently truncated star formation (Dressler
et al.\ 1994; Couch et al.  1994; Stanford, Eisenhardt \& Dickinson
1995).

The above cluster studies have been undertaken alongside similar
HST-based campaigns devoted to the study of the {\em field galaxy
population} (Glazebrook et al.\ 1995; Brinchmann et al.\ 1998; Cohen
et al.\ 2000). Broadly speaking, these studies have likewise
demonstrated an increase in gas content and star formation rate with
redshift although mostly in lower mass systems than those discussed in
the cluster studies. The extent to which the evolutionary trend seen
in the field drives that observed in clusters is thus an interesting
and relevant question.

Much of what has been learned thus far concerning the evolution of
galaxies in clusters has been derived from detailed studies in the
inner $\sim 0.5$ Mpc of several clusters, typically spanning a range
in redshift (Smail et al. 1997; van Dokkum et al. 1998a; Oke, Postman
\& Lubin 1998). Only Couch et al.\ (1998) and van Dokkum et al.\
(1998b) secured morphological data beyond a radius of 1 Mpc via a
WFPC2 mosaic.  Although an efficient strategy in generating a large
sample of spectroscopic members, information is only provided on the
later states of the environmental processes. Important results have
emerged including the relative abundances of elliptical and lenticular
(S0) galaxies as a function of redshift (Dressler et al. 1997; Andreon
et al. 1998; Fabricant et al.\ 2000) and the homogeneity in the colors
of the E/S0 population (Ellis et al.\ 1997; Stanford, Eisenhardt \&
Dickinson 1998; van Dokkum et al.\ 2000).

A major problem that arises in developing a self-consistent
evolutionary picture from observations made with a variety of
clusters is the difficulty of assigning timescales to the
phenomena observed, and connecting clusters at different redshifts
into a temporal sequence. In fact, we have evidence that clusters
evolve continuously by merging and accreting structures from the
surrounding field (e.~g. Zabludoff \& Franx 1993; Abraham et al.
1996a; Balogh et al. 2000) as described in popular theories of
structure formation (e.~g. Kauffmann 1995a,b; Diaferio et al.\
2001). Therefore, we cannot simply assume that a cluster observed
for example at $z\simeq0.6$ evolves ``passively'' to one observed
at $z\simeq0.4$.

Our approach in this series will be to secure high quality data
for a single well-studied system over a large range in
environmental density. A more secure connection with the evolution
observed in field samples may thus be assured and, as we will
demonstrate, the timescale of the star formation activities can be
more readily determined, e.~g. by estimating the likely dynamical
trajectories of infalling galaxies. An important region which
motivated the current study is that where infalling field galaxies
first encounter the cluster potential and the intercluster gas.
Few studies have been devoted to the investigation of this region
either in a statistical sense for an ensemble of clusters (e.g.
Balogh et al.\ 1999; Pimbblet et al.\ 2002) or for individual
clusters (Abraham et al.\ 1996a; Kodama et al.\ 2001).

In a seminal article, Abraham et al.\ (1996b) examined this region via
moderate signal-to-noise ratio spectroscopy to $\sim 5$ Mpc in Abell
2390. An extensive wide-field redshift survey separating field and
cluster members to large radius was central to progress in this
respect enabling the first measures of the radial gradients of
diagnostic spectral features. Unfortunately, the spectroscopic data
was marginal in terms of that required for studies of individual
galaxies, and no HST data was available so a connection with
morphology was not possible. Our series is concerned with developing
and extending the analysis of Abraham et al.\ (1996b) for a single
cluster, Cl0024+16 ($z$=0.4), for which an extensive HST mosaic has
been obtained. The present paper is primarily concerned with
introducing and interpreting the morphological data enabled by our HST
WFPC2 observations in the context of available spectroscopic data.

A plan of the paper follows: in $\S$2 we summarize the scientific
goals which motivate the series and, in turn, define the necessary
observational data.  We then introduce the observational data. In
$\S$3 we describe the strategy for the adopted HST observations, the
data reduction techniques and the production of a catalog that will
serve the entire survey. We define the basis of our morphological
classifications performed to a limiting magnitude\footnote{All
magnitudes are given in the standard Vega system.} of $I=22.5$ and
discuss issues of completeness and bias. This catalog is merged with
the available and new spectroscopic data in $\S$4. 

In preparation for the subsequent analysis of the data, in $\S$5 we
review the various physical mechanisms proposed to explain the
relation between morphology and environment. An important output of
this discussion which guides the later analysis are physical length
scales and timescales associated with the various mechanisms which
define how we will analyze and interpret the data in Cl0024+16, both
as a function of cluster radius and environmental density.  In $\S$6
we analyze the properties of the cluster galaxy population, both as a
function of cluster centric radius and with local projected density,
i.e. the T-R and T-$\Sigma$ relations.  Contrasting the trends seen in
these two relations allows us to interpret the changes in morphology
in the context of specific processes introduced in
$\S$5. Section~\ref{sec:sum} summarizes the results and our main
conclusions.

Throughout the series, we assume the Hubble constant, the matter
density, and the cosmological constant are
H$_0=65$~km\,s$^{-1}$\,Mpc$^{-1}$, $\Omega_{\rm m}=0.3$, and
$\Omega_{\Lambda}=0.7$, respectively. Hence, at the distance of
CL0024+16 ($z=0.395$), $1''$ corresponds to $5.74$ kpc, the bolometric
distance modulus is $DM=41.81$, and the look-back time is 4.6
Gyrs. Finally, $r$ is the radial coordinate in 3-D space, while $R$ is
the radial coordinate in 2-D projected space.

\section{A Wide Field HST Survey of Cl0024+16}

We are undertaking a wide-field HST survey, coupled with suitable
ground-based spectroscopy, for a single cluster at intermediate
redshift. The unique feature is adequate morphological and
spectroscopic data from the core to the turnaround radius where field
galaxies begin their infall. The key scientific goals are: to
determine the mix of galaxy morphologies as function of cluster radius
and local density; to measure the total cluster mass radial profile
beyond the virial radius using weak lensing signals derived from
background galaxies; to determine the star formation properties of
cluster members via diagnostic spectroscopy; to connect these to their
morphological and kinematic properties, and to interactions with the
intercluster medium (ICM); consolidate the above results via a unified
picture for the transformation of the physical properties and
morphologies of galaxies.

Whether such an ambitious list of goals can be uniquely realized via
the detailed study of only a single cluster is of course unclear. Of
necessity, given the challenge of obtaining such a wide body of HST
and ground-based data, it is the logical place to start. Experience
has shown that most astronomical objects/ systems when studied in
considerable detail exhibit unforeseen peculiarities. For example, the
progressive increase in dynamical data for local cluster galaxies has
usually revealed that systems previously considered to be dynamically
relaxed exhibit perceptable substructure (Dressler \& Shectman
1988). Studies based on the statistical properties of member galaxies
as a function of cluster radius must be analyzed with caution in such
circumstances. This would argue for studying several targets and
adopting some form of scaling law for combining data from different
systems (van der Marel et al.\ 2000). On the other hand, at large
radii, it is conceivable that substructures are a dominant feature of
infall patterns which need to be understood rather than averaged out.

Cl0024+16 has a number of features that make it optimal for this
study (c.f. Table~1). There is a wealth of observational data,
including the extensive spectroscopic survey of Czoske et al.
(2001). The choice of an intermediate redshift ($z\approx0.4$) has
many advantages. It ensures the presence of gas rich field
galaxies whose infall is thought to influence the evolution
observed in cluster cores. A large physical area can be observed
with a relatively small number of WFPC2 pointings. Morphologies
and high quality spectra of representative $L^{\ast}$ galaxies can
be obtained in economic exposure times with HST and Keck
respectively. The cluster is also optimally placed along the line
of sight to typical faint ($I\simeq$25) background galaxies for
the measurement and interpretation of weak lensing signals induced
by the cluster. The presence of a multiply-imaged strong lensing
feature of known redshift at the cluster center (Broadhurst et
al.\ 2000) provides an important absolute calibration of the mass
model.

We secured a large wide-field mosaic of HST images for Cl0024+16
through a dedicated program (HST Program 8559: PI Ellis). HST
images were collected from Fall 2000 to Fall 2001 and form the
subject of the first papers of this series, which describe the
measurement of the mass profile of the cluster using the WFPC2 and
STIS images (Kneib et al.\ 2003, in preparation), and the
measurement of the size of the halos of cluster galaxies using
galaxy-galaxy lensing methods (Natarajan et al.\ 2003, in
preparation). Later papers will describe the results of further
Keck spectroscopy of selected members as a means of investigating
the star formation characteristics as a function of cluster
environment.

\section{The WFPC2 Imaging Data}

We now introduce the wide field WFPC2 imaging dataset for
Cl0024+16. After defining the strategy which motivated the HST
observations (\S~\ref{ssec:WFPC2strat}), we describe the data
reduction procedures (\S~\ref{ssec:WFPC2red}) which lead to two
key products: a photometric object catalog
(\S~\ref{ssec:WFPC2cat}) with morphological classification
(\S~\ref{ssec:WFPC2morph}), which we will refer to as the WFPC2
catalog.

\subsection{Observing Strategy}

\label{ssec:WFPC2strat}

In order to probe both the transition and peripheral regions of the
cluster (see $\S$~\ref{ssec:2D}), a sparse-sampled pattern of WFPC2
fields was necessary. We defined a circle of diameter $\sim 25'$
($\sim$10 Mpc) and arranged 38 non-contiguous pointings which,
together with earlier archival data, fully cover the center and sample
the periphery at $\simeq$30\% area coverage. This strategy ensures we
cover the required range of environments and cluster radius for the
galaxy population studies. The arrangement was also chosen to permit
azimuthal-averaging of fields for radial trends, e.g. for the weak
shear profile (c.f. Kneib et al.\ 2003 for a discussion of the
observational strategy as defined from the lensing goals).

\begin{figure*}
\begin{center}
\resizebox{\textwidth}{!}{\includegraphics{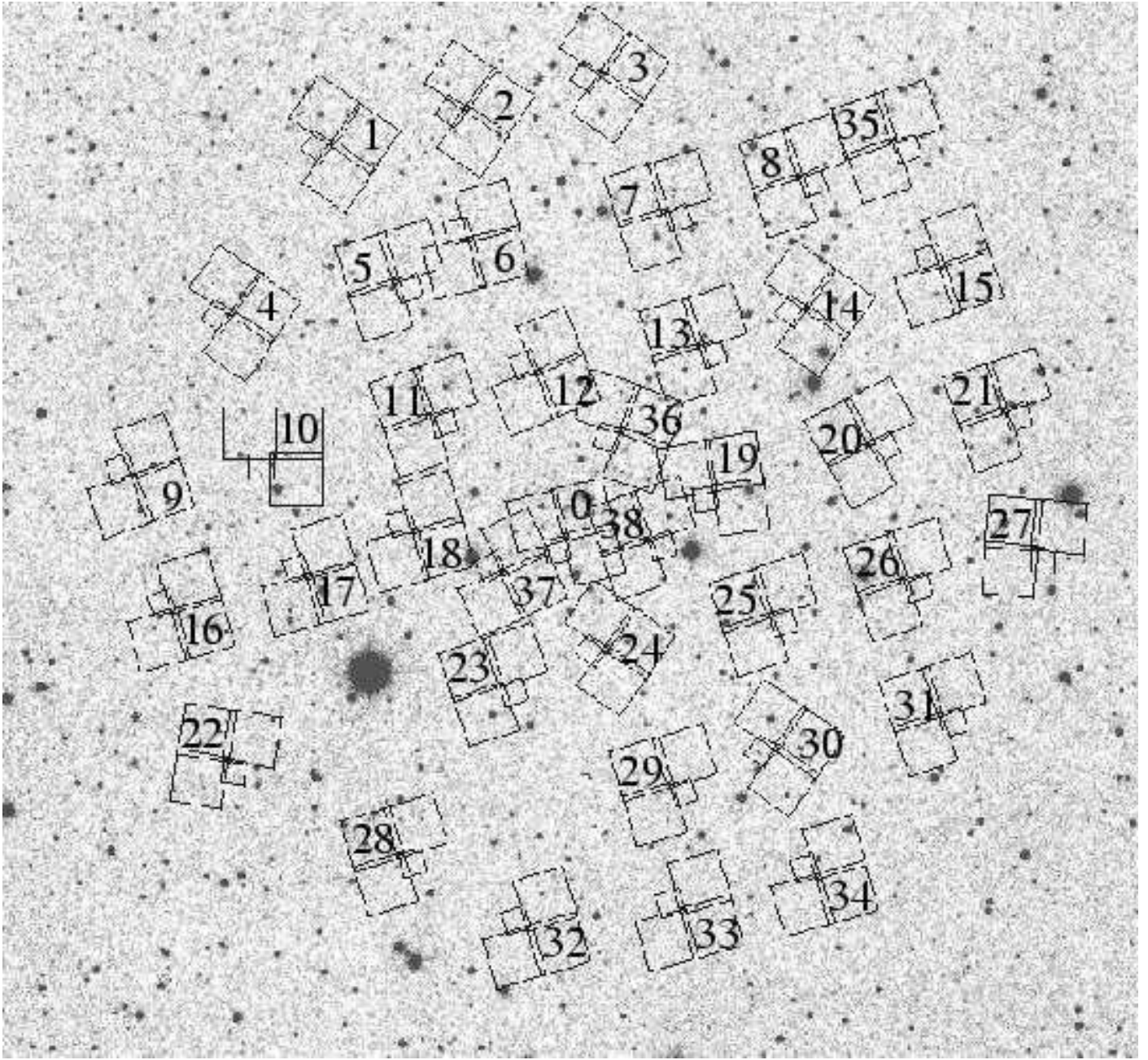}}
\end{center}
\figcaption{The arrangement of WFPC2 fields observed in Cl0024+16.
The WFPC2 footprints are overlaid on a Digital Sky Survey image of
the field (N is up, E is left). Each WFPC2 field is approximately
$80''$ on a side. Details of each pointing are given in
Table~\ref{tab:pointings}.
\label{fig:map}}
\end{figure*}

A list of pointings, coordinates and orientations is given in
Table~\ref{tab:pointings} and the field distribution is portrayed in
Fig.~\ref{fig:map}. In the original approved schedule, the WFPC2 field
positions and orientations were carefully chosen to ensure some
overlap of parallel STIS observations. The higher spatial resolution
of STIS can be used to great advantage in calibrating the WFPC2 point
spread function for weak lensing studies. Unfortunately, for
scheduling reasons, this carefully constructed plan had to be
abandoned, and, in order to safeguard completing the survey prior to
the launch of ACS, we had to discard all orientation restrictions on
the WFPC2 pointings soon after observations commenced.

All WFPC2 images were taken in the F814W filter providing photometry
for cluster members which approximates the rest-frame V-band
(hereafter, for simplicity, we will refer to F814W as $I$).  The
filter choice was motivated by the need for an adequately faint sample
of background galaxies for the weak lensing analyses, while ensuring
adequate comparisons with existing and ongoing studies of intermediate
redshift galaxies (e.g. Dressler et al. 1997; Brinchmann et al.\
1998). At each position, four equal exposures totaling two HST orbits
(4000s-4400s, see Tab~\ref{tab:pointings}) were dithered with a
$2\times2$ grid with half-integer offsets. For the central field
(POS00), we utilized the image originally analyzed by Colley et
al. (1996) referred to in the Morphs collaboration (Smail et al.\
1997) (GO: 5453, PI: Turner) whose exposure time was 18,000s in F814W
and 25,200s in F450W.

\subsection{Data Reduction}

\label{ssec:WFPC2red}

We implemented a pipeline based on the {\sc iraf} package {\sc dither}
(Fruchter \& Hook 2002) to reduce the WFPC2 data. The pipeline cleans
and combines the images, by iteratively refining the measurement of
relative offsets and the cosmic rays/defect identification (the
algorithm is similar to the one described by Fruchter \& Mutchler
1998). The output of the pipeline consists of two sets of images, one
with a pixel size equal to the original pixels size ($0\farcs1$ for
the WF chips), and one with pixel size half the original ($0\farcs05$
for PC chips). An exposure time map is produced for each chip.

The quality of the combined images was carefully checked by visual
inspection of all the intermediate steps of the pipeline and by a
variety of quality checks including requirements on the homogeneity of
the exposure time maps to ensure homogeneous photometry, and on the
sharpness of the point spread function (the Gaussian-FWHM of stars as
measured with {\sc imexam} was generally slightly better in the
combined images than in the original ones, due to better sampling).

\subsection{The Photometric Catalog}

\label{ssec:WFPC2cat}

The photometric catalog derived from the WFPC2 images will be used
for most subsequent applications in the series of papers, from the
morphological classification discussed later in this paper, to the
identification of background samples for weak lensing studies
(Kneib et al.\ 2003; Natarajan et al.\ 2003), and later for target
selections for the spectroscopic campaigns. A reliable catalog
with few spurious detections and well-understood completeness
characteristics is essential. A high degree of completeness at the
faintest limits ($I\ga 25$) is less important than a
well-understood completeness function because weak lensing signals
depend on the angular size of a faint galaxy and cannot uniformly
be detected in faint galaxies of a given apparent magnitude (e.~g.
Bacon, Refregier \& Ellis 2000).

Faint object identifications were performed using the Source Extractor
algorithm SExtractor (Bertin \& Arnouts 1996), one of the standard
packages for this purpose (e.~g., Casertano et al.\ 2000). After some
experimentation, it was decided to adopt detection criteria similar to
those used by the Morphs collaboration\footnote{$0.12 \square''$ above
24.6 mag arcsec$^{-2}$ for the WF chips, $0.09 \square''$ above 23.6
mag arcsec$^{-2}$ for the PC.  Photometric zero points were taken from
the HST Data Handbook.}. The deblending threshold and contrast
parameters were set to 32 and 0.03 respectively, ensuring nearby
sources are separated while fainter objects that appear part of the
same system (e.~g., bright knots on spiral arms) are preserved as a
single entity. These choices affect completeness somewhat at the
faintest magnitudes -- $I\ga25$ in our case -- (c.f. discussion in
Casertano et al.\ 2000), but will not affect the results discussed in
this paper, which are mostly concerned with the brighter galaxies. The
catalog does not include the area of reduced sensitivity covered by
the PC which represents only a modest fraction of the total area.

A basic measure of catalog completeness can be determined from the
I-band galaxy number counts after excluding duplications from
overlapping fields (Fig.~\ref{fig:numcou}). Although obviously a
clustered field, excluding the data from the central POS00, the
Cl0024+16 counts are only modestly above those of genuine field
samples, mostly at $I\sim20-22$; this gives a measure of the challenge
of locating cluster members at large radii. The turnover fainter than
$I\sim25$ serves as an indicator of the limiting depth of our
catalog. As expected, the central POS00 field (red stars) shows a more
significant excess at magnitudes fainter than the brightest cluster
galaxy (BCG; $I\sim 18$; see also Smail et al.\ 1997). The faster
turnover of the central POS00 field at $I\ga25$ is the result of
crowding, magnification bias (Fort et al. 1997; Dye et al. 2002), and
possibly cosmic variance (see, e.g., Casertano et al.\ 2000).

As some fields overlap (see Fig.~\ref{fig:map}), we have an
opportunity to verify photometric consistency. No systematic
differences were found and the scatter is less than 0.05 mags rms
at $I<22.5$ increasing to $\sim 0.3-0.4$ mags at $I\sim25$ rms. As
expected, this rms scatter is slightly larger than the random
error (0.2-0.3 mags) due to the additional systematic components
in the detection algorithm and in the computation of the total
magnitude (mag\_auto, see Bertin \& Arnouts 1996). A $\sim
3\sigma$ detection at $I\sim25$ is consistent with the observed
turnover in the number counts. All HST images were registered to
the astrometry of Czoske et al.\ 2001 (hereafter the CFHT
astrometry), which has been determined to be accurate to
$\la0\farcs2$ from comparisons with the USNO-A2.0 catalog. Offsets
between the original WFPC2 astrometry and the CFHT astrometry are
typically $\simeq0\farcs5$, and in the overlap regions we find an
rms scatter of $\sim0\farcs3$, consistent with the estimated
uncertainty in the CFHT astrometry.

\begin{inlinefigure}
\begin{center}
\resizebox{\textwidth}{!}{\includegraphics{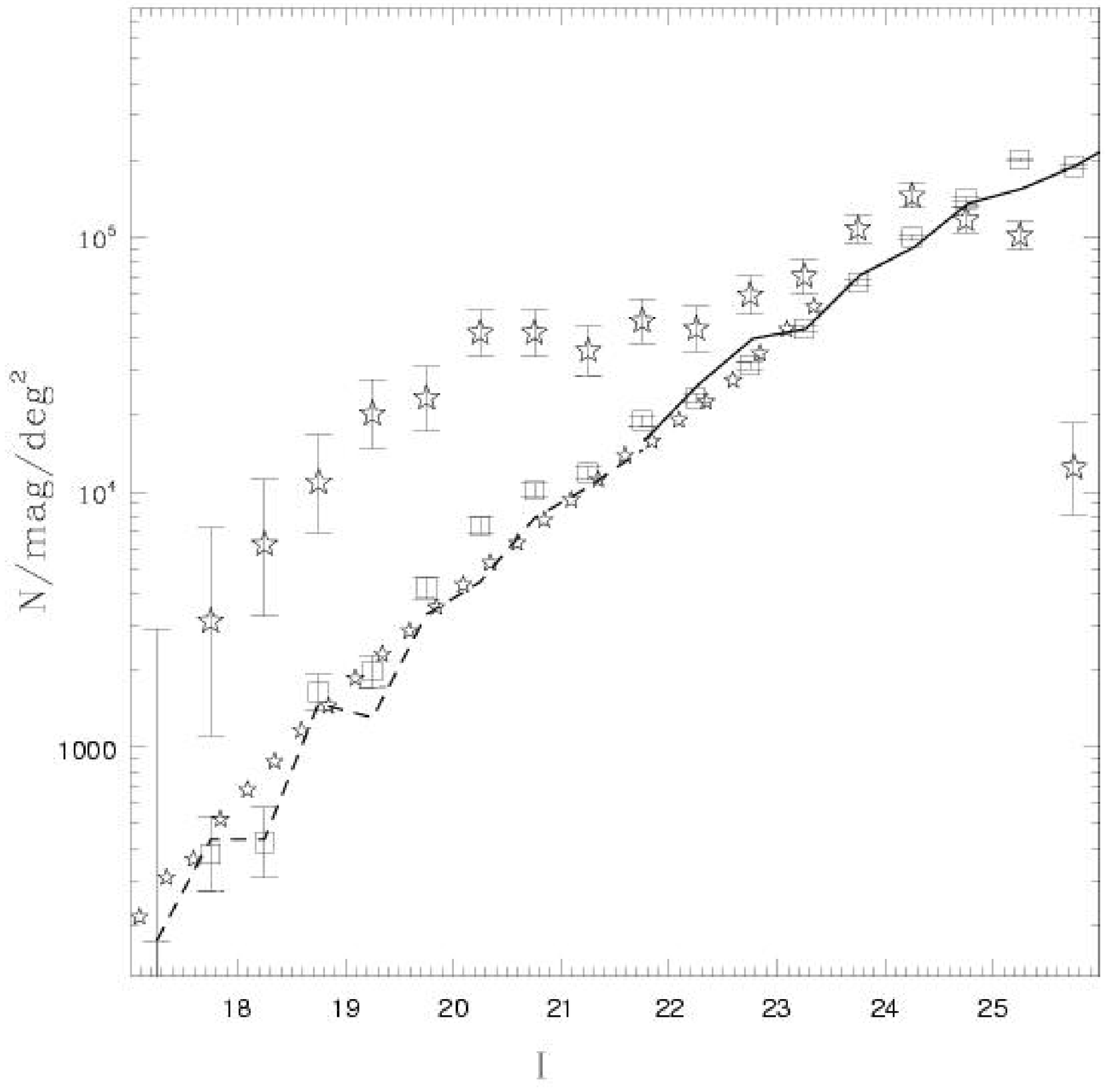}}
\end{center}
\figcaption{I-band number counts from the Cl0024+16 photometric
catalog (open red stars: central POS00 field; open black squares: remaining
area) compared with data from the Hubble Deep Fields (solid line,
Williams et al.\ 1996, Casertano et al.\ 2000) catalog, Postman et
al.\ (1998; open stars) and the HST-Medium Deep Survey catalog by
Abraham et al.\ (1996; dashed line). The turnover at I$\simeq$25.5
indicates where the catalog becomes significantly incomplete. \label{fig:numcou}}
\end{inlinefigure}

The photometric catalog comprises 22,000 objects to
$I\simeq$25\footnote{The catalog is available electronically at
\rm{http://www.astro.caltech.edu/~tt/0024}}.  Entries are described in
Tab~\ref{tab:entries} and the head of the catalog is given in
Tab~\ref{tab:catalog} In the central region, independent pointings
partially overlap, and duplications have not been removed. All
magnitudes are in the Vega system and corrected for galactic
extinction, adopting E(B-V)=0.057 (Schlegel et al.\ 1998).

\subsection{Morphological classification}

\label{ssec:WFPC2morph}

We now discuss the morphological classification of the objects in
the catalog. Broadly, the procedure followed that adopted for the
HST {\em Medium Deep Survey} (Abraham et al.\ 1996a) for which the
minimum typical exposure time (4000s) was similar and a limiting
magnitude of $I$=22.0 was adopted. After some experimenting by one
of the classifiers (RSE) in that study, it was determined that a
similar degree of reliability would be achieved in the Cl0024+16
fields to a fainter limit $I$=22.5. This value is intermediate
with the fainter limit ($I$=23) adopted in the 6300-16,800 sec
Morphs study (Smail et al.\ 1997).

As noticed in previous studies, the uncertainty in the
classification is a strong function of signal-to-noise ratio close
to the limit and some distinctions are more critical then others,
e.~g. distinguishing amongst regular galaxies is more reliable
than disentangling late types from merger and irregulars. Thus,
after describing our classification procedure, we discuss in
detail the uncertainty (\S~\ref{ssec:unmorph}).  A summary is
given in~\S~\ref{ssec:summorph}.

\begin{figure*}
\begin{center}
\resizebox{\textwidth}{!}{\includegraphics{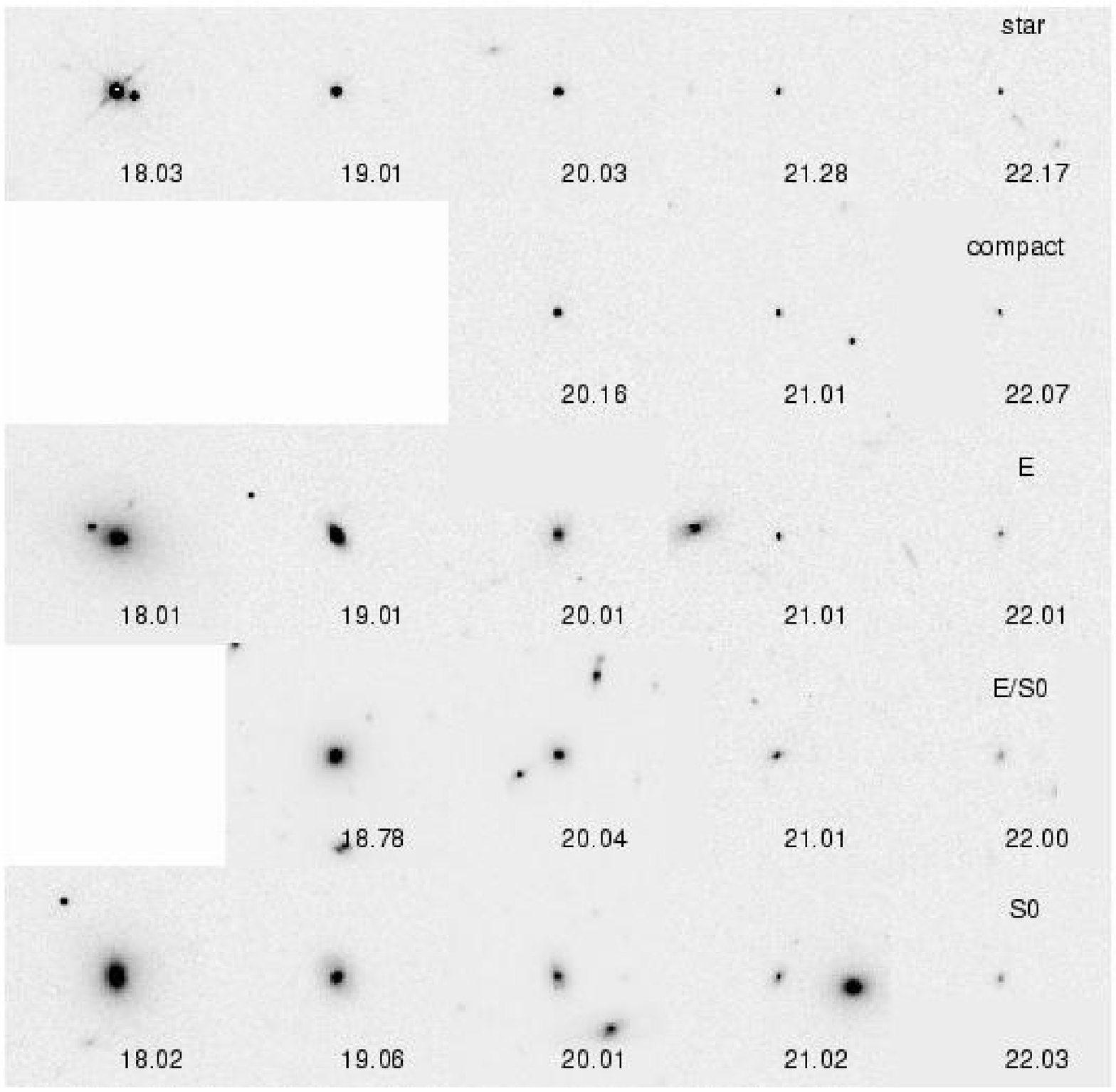}}
\end{center}
\figcaption{Montage of galaxies (each $12\farcs8$ on a side) in
the Cl0024+16 field sorted by type and magnitude. In each row we
show typical galaxies for a given morphology, from $I\sim18$ to
$I\sim22$ (left to right) in steps of $\sim1$ magnitude. Empty
spaces indicate no galaxy of that type is available in the
respective magnitude bin. Magnitudes are indicated at the lower
right corner of each image, morphological types at the top-right
corner of each row. The display scale is linear, with uniform
stretch levels.
\label{fig:fig_morph1}}
\end{figure*}

\begin{figure*}
\begin{center}
\resizebox{\textwidth}{!}{\includegraphics{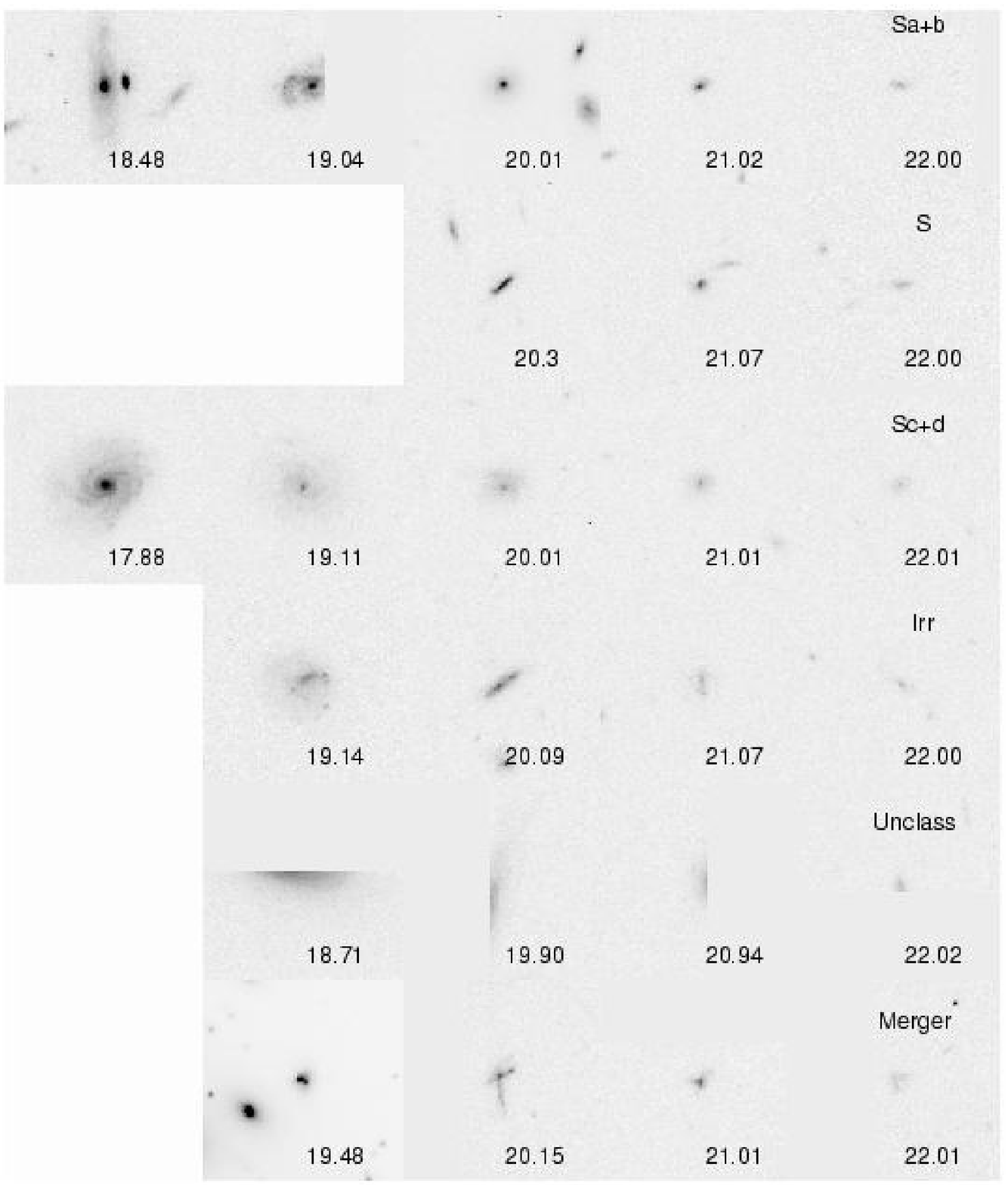}}
\end{center}
\figcaption{As Figure~\ref{fig:fig_morph1} for morphological types
Sa+b (3) to Mergers (8).
\label{fig:fig_morph2}}
\end{figure*}

Morphological classification of all objects with magnitude (mag\_auto)
brighter than $I=22.5$ was performed by one of us (RSE) using an {\sc
iraf} task which displayed, for each entry, a square region
$12\farcs8$ on a side, on a logarithmic intensity scale to simulate
appearance on a photographic plate. Each object was simultaneously
displayed at four stretch levels to ensure an adequate dynamic
range. Morphologies (hereafter T) were assigned according to the MDS
scheme introduced by Abraham et al.\ (1996a): -2=star, -1=compact,
0=E, 1=E/S0, 2=S0, 3=Sa+b, 4=S, 5=Sc+d, 6=Irr, 7=Unclass, 8=Merger,
9=Fault. Examples for classes -2 to 8 are shown in
Figures~\ref{fig:fig_morph1}, and~\ref{fig:fig_morph2}. It is
important to clarify that the nomenclatures Sa+b and Sc+d represent the
union of subclasses, not intermediate spiral types. As in the MDS, the
system is hierarchical: classes such as E/S0 or S are used when the
classifier recognizes the galaxy as an early or late type, but no
further subclassification can be determined. The Unclass (7) category
comprises objects that are only partially imaged, e.g.  they fall on
the edge of a chip. Finally, T=8 (Merger) is used when the
isophotes of two objects of comparable brightness merge (c.f. examples
in Fig~\ref{fig:fig_morph2}). This could, of course, in some cases be
due to projected alignments.

The classifier's comments were also recorded together with the
morphological type. A total of 2181 objects were processed in this
way, including the ones in the central field classified by the Morphs
collaboration. The distribution of morphological types as a function
of magnitude (without regard to cluster membership) is shown in
Figure~\ref{fig:tfrac}.

\subsubsection{Uncertainties in the Morphological Classifications}

\label{ssec:unmorph}

Internal consistency in our classifications can be obtained by
examining the overlapping fields, given our {\sc iraf} task ensures a
blind treatment of each image.  At $I<21$: 48 \% of the galaxies have
identical morphological type, 89 \% differ by at most $\Delta T=1$.  At
fainter magnitudes ($21<I<22.5$) 56 \% of the galaxies have identical
T, while 74 \% differ by at most $\Delta T=1$. No clear magnitude
dependence is seen. If we consider only ``broad classes", defined as
stars (T=-2), compacts (-1), early-types (E+S0, 0 to 2), spirals (3 to
5), and Irr/Unclass/Merger (6 to 8) galaxies, the consistency improves
significantly. 75-80 \% of galaxies have the same broad class and {\em
all} lie within $\pm$ 1 broad class.  The agreement is consistent with
earlier similar studies (e.~g.  Smail et al.\ 1997; Fabricant et al.\
2000).

A further step was undertaken to ensure internal consistency.
Postage stamps $12\farcs8$ on a side were sorted by morphological
type and magnitude and carefully examined by two of us (RSE and
TT). By comparing images of comparable signal-to-noise, the
consistency and repeatability of the classification was
re-examined. Types were only changed in $<10$\% of the galaxies,
mostly within broad classes (e.g. Sa+b to S).

Two of the well-known critical points in the morphological
classification, are E+S0 vs spiral separation, and star galaxy
separation (see e.~g. Smail et al.\ 1997; Fabricant et al.\ 2000).
The difficulty of separating E+S0 and spirals has been discussed
widely in the literature. Unfortunately -- as opposed to the star
galaxy separation where spectroscopic information provides an
unambiguous verification -- using external information to verify
E+S0 vs spirals morphologies is much more difficult. Although
certain properties, such as spectral features and colors, are
known to correlate with morphology there is significant scatter,
and, as we are interested in the stellar population properties of
galaxies, their use could lead to significant biases when applied
to evolutionary studies.

Distinguishing features (spiral arms, bright young stars and star
forming regions) may be lost at intermediate redshift because of
surface brightness dimming ($\propto (1+z)^{-4}$) and resolution
and/or sampling effects. Figures~\ref{fig:fig_morph1}
and~\ref{fig:fig_morph2} show that bright E+S0 and Sa+b galaxies can
be separated brighter than $I\la20$. Our classifications are
internally consistent to $I\sim21$ but suffer some ambiguity at
$I=21-22.5$. As an external check, the study by Fabricant et al. \
(2000) of Cl1358+62 is particularly relevant since the HST
observational setup (instrument, filter, and exposure time) and
cluster redshift are closely similar to those for Cl0024+16. Fabricant
et al. found that human classifiers consistently distinguish
early-types and spirals to $I\sim21$, with increasing uncertainty
beyond this limit. Finer classification was found to be much more
uncertain. For example the two classifier groups in Fabricant et al.\
(2000) yielded discrepant values of the ratio of S0 and E galaxies
already at $I<21$. We conclude that early-type and spiral galaxies can
be consistently distinguished with our instrumental setup down to at
least $I\sim21$.

Two further effects should be considered. Firstly, if the
distribution of morphological types is not uniform, any scatter
might smear the distribution by transferring objects from the most
populated class to adjacent classes (see e.g. discussion in the
Appendix of Fabricant et al.\ 2000). The second is due to the
specific effects of distance, i.e. loss of resolution and surface
brightness dimming. Qualitatively, these effects tend to suppress
low surface brightness features (such as extended envelopes around
S0 galaxies) and smear out small-scale features (such as compact
star forming regions, or spiral arms). Thus, the net qualitative
effect is to transfer S0 galaxies to the E class and Sa+b galaxies
to the S0 or E class. Unfortunately, the magnitude of these biases
depend strongly on the properties of the parent populations and
therefore cannot be accurately corrected without additional
assumptions. However, in our observing conditions, these effects
are likely to be small especially at $I\la21$ (see, e.g.,
discussion in Glazebrook et al.\ 1995; Abraham et al.\ 1996a;
Ellis et al.\ 1997).

Securing a reliable distinction between stars, compact galaxies,
and faint ellipticals is also an important concern. At bright
magnitudes ($I\la20$) distinguishing between these types is
straightforward (Figure~\ref{fig:fig_morph1}). Fainter than
$I\sim20$, diffraction spikes cannot be seen and peaked nuclear
emission might be present in some galaxies. Classification becomes
troublesome for objects smaller than the WFPC2 F814W PSF whose
FWHM$\approx0\farcs15$ (Casertano et al.\ 2000), corresponding to
approximately $0.9$ kpc at the cluster redshift. Fortunately, most
normal elliptical members at $z\simeq0.4$ within our magnitude
limit should have effective radii significantly larger than 0.5
kpc (Ziegler et al.\ 1999 and references therein), and therefore
should be clearly resolved.

Our spectroscopic catalog (see $\S$~\ref{sec:z}) provides a valuable
external check on the star compact elliptical boundary: out of 13
objects classified as stars with available spectroscopic data, 11 are
confirmed and 2 have $z\sim0.3$ flagged with ``uncertain" quality.
Out of 8 objects with spectra classified as compacts (i.e. deduced
morphologically to be non-stellar), 5 are stars and 3 are galaxies
with $z\approx0.3-0.7$. Although the samples are small, the
spectroscopic data suggests we have erred on the side of trying not to
omit any extragalactic sources. This augurs well and suggests no
ellipticals have been misclassified as stars. Indeed, all of the objects
with zero spectroscopic redshift were independently morphologically
classified as stars or compacts.

\begin{inlinefigure}
\begin{center}
\resizebox{\textwidth}{!}{\includegraphics{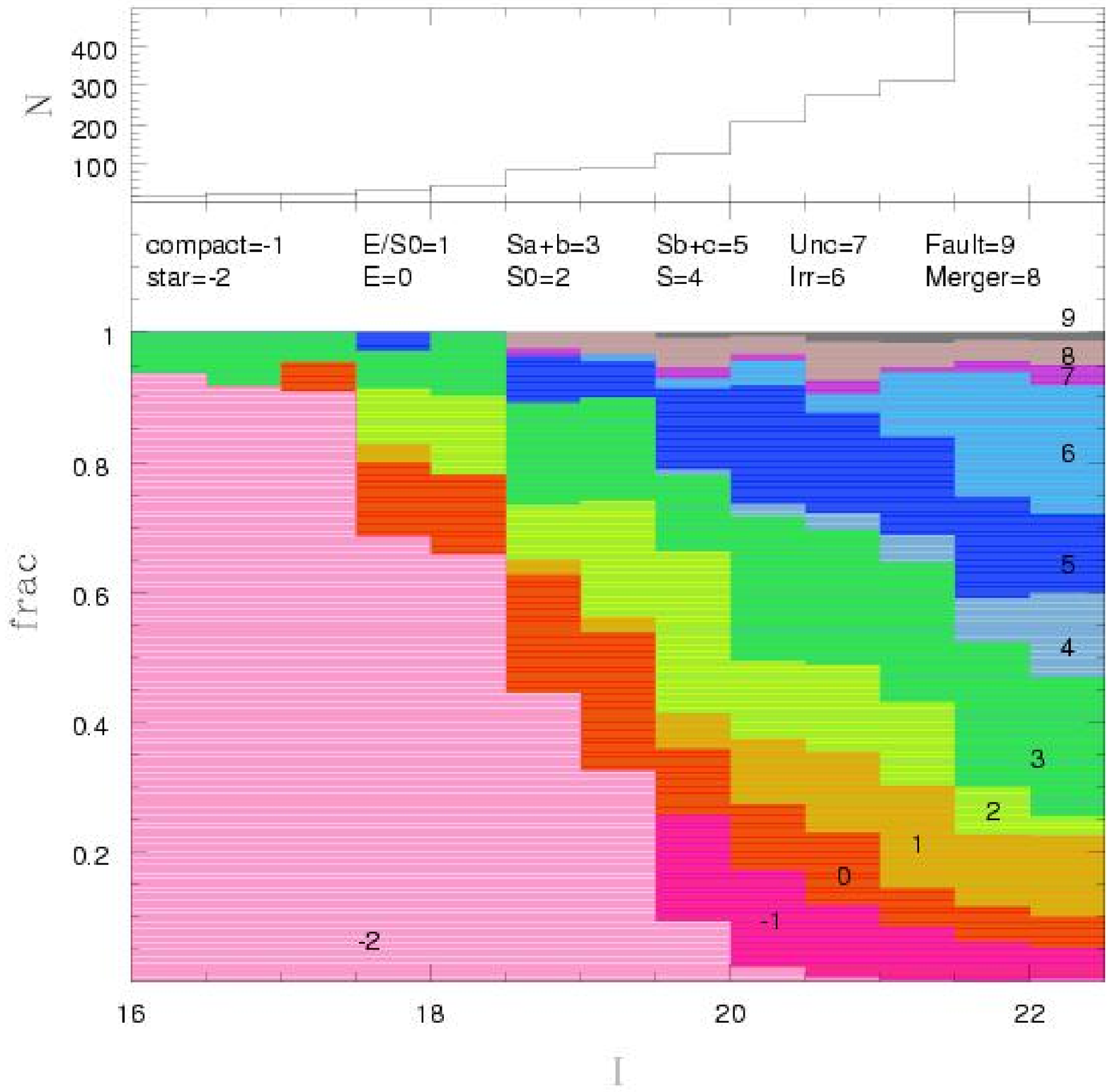}}
\end{center}
\figcaption{The distribution of morphological types as a function of
$I$ magnitude in the field of Cl0024+16; no distinction is made
between members and non-members.
\label{fig:tfrac}}
\end{inlinefigure}

A second external check is facilitated by considering the
photometric distribution of objects in the color-magnitude
diagram. For this purpose we augmented the HST photometry with
wide field V and I band images obtained with Canada France Hawaii
(CFHT) 12K camera. The E+S0 members define a tight color-magnitude
sequence (see Natarajan et al.\ 2003). Stars span a wider range,
generally bluer then the E+S0 red-sequence, and are found at
significantly brighter magnitudes. Compacts are numerous only at
$I\ga20$, and also span a wide range of colors although are
generally redder than the E+S0 red sequence, indicating that some
are likely to be background galaxies.

A final check is provided by the SExtractor parameter CLASS\_STAR
(see Bertin \& Arnouts 1996 for details), which should be close to
unity for genuine stellar objects and around 0 for extended
objects. Confirming the results of our discussion, non-saturated
stars have CLASS\_STAR parameters greater than 0.9, while E
galaxies have typically CLASS\_STAR close to zero, except for a
small fraction at magnitudes fainter than $I\sim21$ that have
CLASS\_STAR larger than 0.8. Compacts have CLASS\_STAR parameters
between 0.8 and 1. In summary therefore, external diagnostic
information based on spectroscopy and colors support the notion
that there is little or no confusion so far as our morphological
selection of cluster spheroidals, and virtually all of the
compacts are either stars or distant background galaxies.

\subsubsection{Summary of the morphological catalog}

\label{ssec:summorph}

We have morphologically classified 2181 objects in the
SExtractor catalog to a magnitude limit of $I<22.5$, using the
morphological classes illustrated in Figures~\ref{fig:fig_morph1}
and~\ref{fig:fig_morph2}. The resulting distribution of types as a
function of magnitude is shown in Figure~\ref{fig:tfrac}.  The
accuracy of the classification -- measured in terms of repeatability
-- is found to be excellent down to $I\sim21$, and still very good in
broader classes (star; compact; E+S0; Sp; Irr/merger) to the limit
$I=22.5$, in agreement with the precision attained independently in
similar datasets (Fabricant et al.\ 2000).  Two critical points such as
star/galaxy separation and S0 vs Sa separation are investigated with
the aid of external information. We find that all of the
spectroscopically confirmed stars are classified either as star or as
compact, a class that is used mostly at magnitude fainter than
$I\sim20$, and is most likely a mix of stars, and high redshift
compact objects. Similarly we conclude that the quality of the data is
sufficient to separate reliably early-type and late-type galaxies down
to at least $I\sim21$.  The subsequent analysis based on morphological
types will note the distinction in quality beyond $I\sim$21 and often
be restricted to this limit where the classifications are particularly
reliable.

\section{The Spectroscopic Data}

\label{sec:z}

Following the comprehensive spectroscopic survey of Czoske et al
(2001), a sufficiently large sample of spectroscopically confirmed
members is already available to support the present investigation
of the distribution of morphological types with radius and local
density.

A spectroscopic survey of Cl0024+16 is continuing at the 10m Keck
telescopes with two goals: {\it i)} to obtain high-quality and where
possible resolved 2-D spectra in order to explore dynamical and star
formation characteristics for members of known morphology; {\it ii)}
to extend the redshift survey beyond the limit of $I\sim21$ of
previous surveys (Dressler et al.\ 1999; Czoske et al.\ 2001) and gain
a more complete sampling that is better matched to the dilute pattern
of WFPC2 images. Although this survey will continue, we briefly
present here the spectroscopic catalog incorporating all measures in
the literature or made available to us, plus the initial results from
our Keck campaign\footnote{The complete redshift catalog will be
released at the end of the campaign.}.

The strategy, observations and analysis are discussed in
Sec~\ref{ssec:keck}. The catalog is discussed in
Sec~\ref{ssec:speccat}. A separate discussion of the completeness of
the catalog is given in Sec~\ref{ssec:speccomp}, bearing in mind the
requirements for Section 6.

\subsection{Keck Spectroscopy: strategy, observations \& analyses}

\label{ssec:keck}

The new Keck spectra discussed here were obtained using the Low
Resolution Imager Spectrograph (LRIS; Oke et al.\ 1995). In order to
cover a suitable range of local density we decided to survey a strip
oriented approximately North-South through the cluster center. The
orientation was chosen to maximize overlap with WFPC2 fields with
known cluster members.

Spectroscopic targets were selected from the CFHT I-band mosaic
(Czoske et al. 2003, in preparation) according to the following
priority: {\it i)} known members with HST morphology; {\it ii)}
galaxies from the WFPC2 catalog without redshifts in the range
21$<I<$22.5; {\it iii)} known members without WFPC2 imaging; {\it iv)}
objects within 21$<I<$22.5 without WFPC2 imaging. In the case of
spectroscopically-confirmed members, where possible tilted slits were
cut for those classified as edge-on spirals in order to offer the
possibility of securing rotation curves from extended line emission. A
detailed study of resolved spectroscopy will be deferred until a later
paper, pending completion of the spectroscopic campaign. In this paper
we use the newly determined redshifts to update the list of members.

Keck I LRIS observations were conducted on Oct 17-19 2001. For the red
arm we mounted the 600/5000 grating providing wavelength coverage 5000
\AA\, to 7500 \AA\, with a pixel scale of approximately $1.25 \AA
\times 0\farcs215$. The blue arm covered the region bluewards of 5100
\AA\ with the 600/4000 grating and a pixel scale of approximately
$1\AA \times 0\farcs215$.  Multi-object masks were milled with
slitlets $1''$ wide providing spectral resolutions of $\sigma \sim 90$
kms$^{-1}$ (red) and $\sigma \sim 130$ \kms (blue). Five masks were
exposed yielding new data for 107 slitlets.  The exposure time was
$4\times1800s$ and the seeing was $0\farcs8$ - $1\farcs2$ FWHM;
conditions were generally non-photometric.  After each set of four
science exposures, internal flat field and Hg-Kr-Ne-Ar lamps were
obtained without moving the telescope to obtain accurate fringing
removal and wavelength calibration.

After bias removal the individual slitlets were separated and reduced
in a standard manner (see e.g. Treu et al.\ 1999, 2001a). Extracted
spectra were inspected by two of us (TT and RSE) and 92 redshifts
measured by identifying absorption and emission lines (when
present). A quality flag -- in a scheme similar to that of Czoske et
al.\ (2001) -- was assigned to each spectrum: 1=secure (63 objects),
2=probable (12), 3=possible/uncertain (17). The latter category mostly
includes spectra where only a single feature can be detected.

\subsection{The Redshift Catalog}
\label{ssec:speccat}

The Cl0024+16 redshift catalog is based on the new Keck data, the
compilation from Czoske et al.\ 2001 (which includes Dressler et
al.\ 1999 observations) and unpublished redshifts kindly provided
by F.~Owen (2001, private communication) and A.~Metevier \& D.~Koo
(2001, private communication). In total 1018 redshifts, including
duplicates, are available, independently of the WFPC2 data.
Duplications arise in several ways. For some galaxies, redshifts
are available from more than one source. The typically agree to
within 0.001; we adopt the average and the rms scatter as the
uncertainty. Discrepant redshifts were found for only two galaxies
which were removed from the combined catalog.

The final catalog comprises redshifts for 787 objects. As noted by
Czoske et al.\ (2002), the distribution around $z\sim0.4$ is
characterized by a main peak (A in Fig.~\ref{fig:histoz}), a secondary
peak at lower redshifts (B), interpreted by Czoske et al.\ (2002) as a
foreground system, and a third sparser peak of galaxies at higher
redshift. For a thorough discussion of this complex redshift
distribution and possible interpretations see Czoske et al.\
(2002). Here we only remind that the relative velocity between the two
peaks is $\sim 3000$ \kms\, in the rest frame of the cluster, and that
the formal velocity dispersions of the two peaks are approximately 600
\kms. Adopting the velocity limits for the peaks as defined by Czoske
et al.\ (2002), there are 304 galaxies in peak A ($0.387<z\le0.402$)
and 54 in peak B ($0.374<z\le0.387$). Matching the above catalog with
the WFPC2 catalog yields what we will term the WFPC2$-z$ catalog
containing 362 objects, including 15 stars.  Of the galaxies with
WFPC2 morphologies, 179 have redshifts within peak A but only 16 lie
within peak B (Fig.~\ref{fig:histoz}). In the following we will use
the term cluster members to refer to galaxies in either one of the two
peaks (the results are unchanged if only peak A is considered).

\begin{inlinefigure}
\begin{center}
\resizebox{\textwidth}{!}{\includegraphics{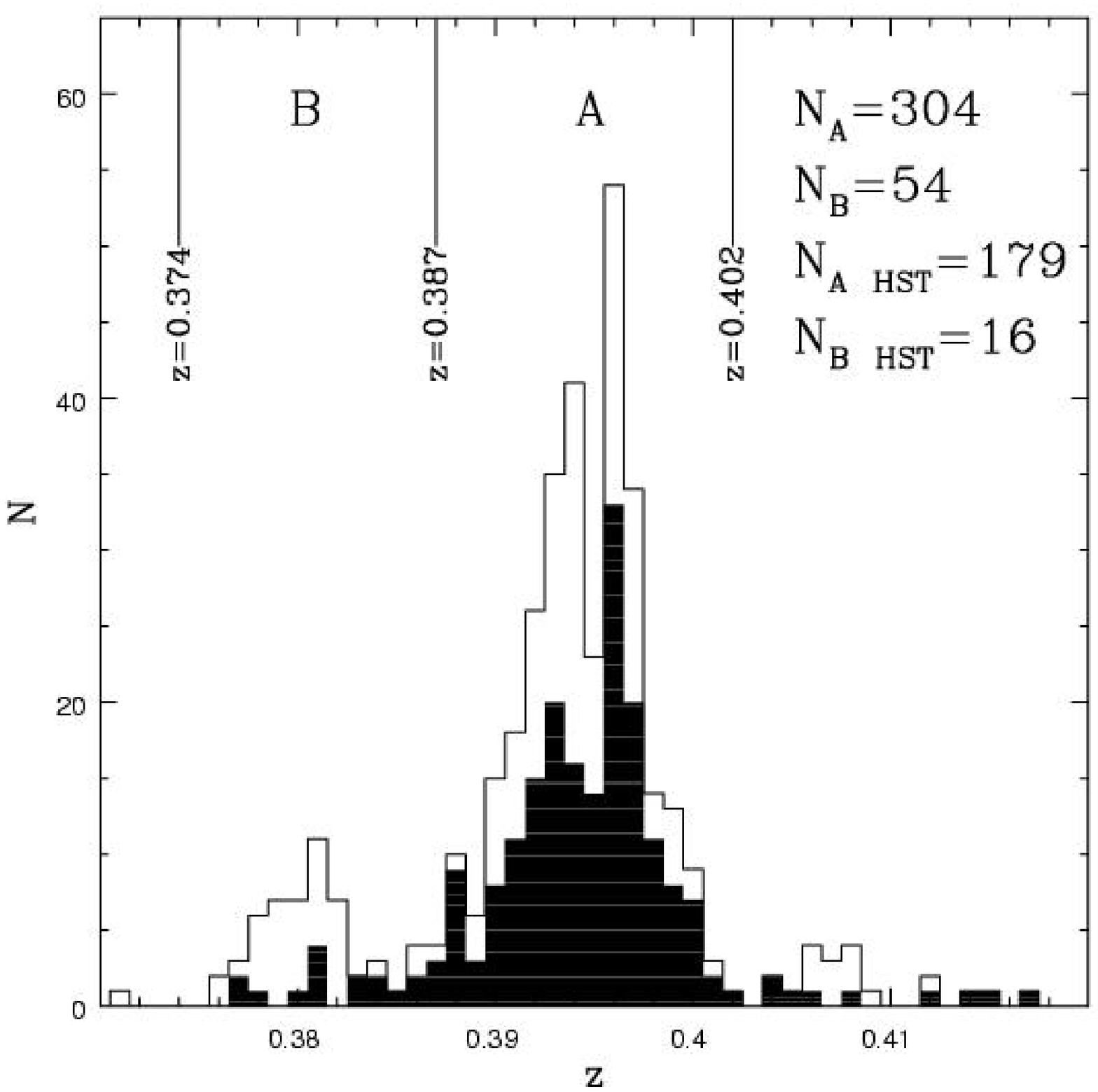}}
\end{center}
\figcaption{Redshift distribution of galaxies in the field of
Cl0024+16 in the vicinity of $z\sim0.4$.  The distribution from the
parent redshift catalog (Section~\ref{ssec:speccat}) is shown as an
empty histogram, while that from the WFPC2-z catalog is shown as a
filled histogram. The double peak in the redshift distribution is
discussed by Czoske et al.\ (2002). \label{fig:histoz}}
\end{inlinefigure}

Figure~\ref{fig:spacedistAB} shows the projected distribution of
members in peaks A and B. As noticed by Czoske et al.\ (2002), those
in peak B are fairly homogeneously distributed. By contrast, there is
a marked clump of those in peak A to the NW of the cluster center.
Because the fraction of area covered by our WFPC2 survey decreases
with radius and the A peak is more concentrated than the B peak -- the
ratio of galaxies in peak B to galaxies in peak A is smaller in the
WFPC2-z sample than in the full spectroscopic sample.

\begin{inlinefigure}
\begin{center}
\resizebox{\textwidth}{!}{\includegraphics{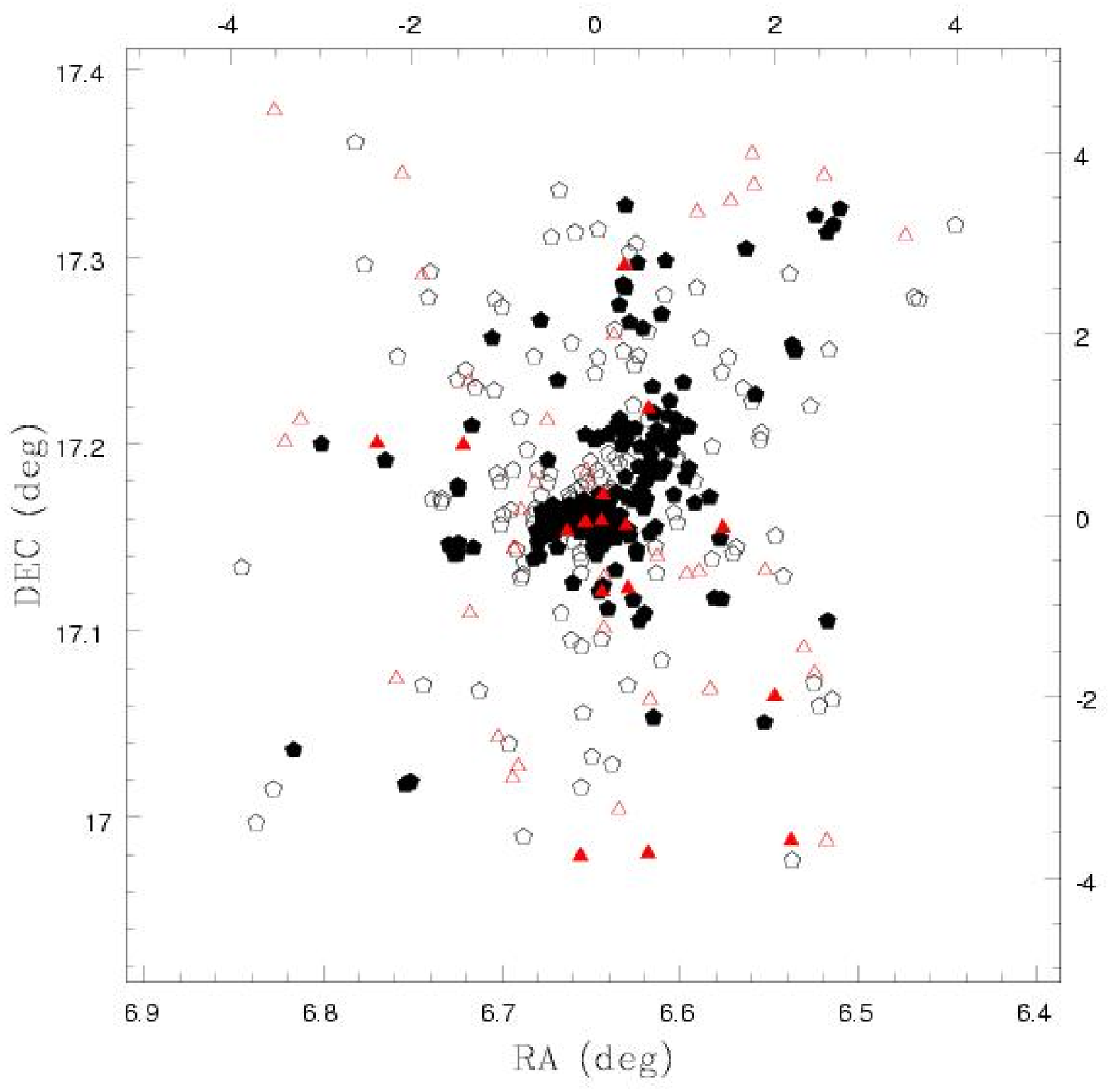}}
\end{center}
\figcaption{Projected distribution of Cl0024+16 member galaxies (see
text for details) in redshift peaks A (black pentagons) and B (red
triangles). Solid symbols indicate objects with available HST-WFPC2
images. The top and right scale is in Mpc. \label{fig:spacedistAB}}
\end{inlinefigure}

\begin{inlinefigure}
\begin{center}
\resizebox{\textwidth}{!}{\includegraphics{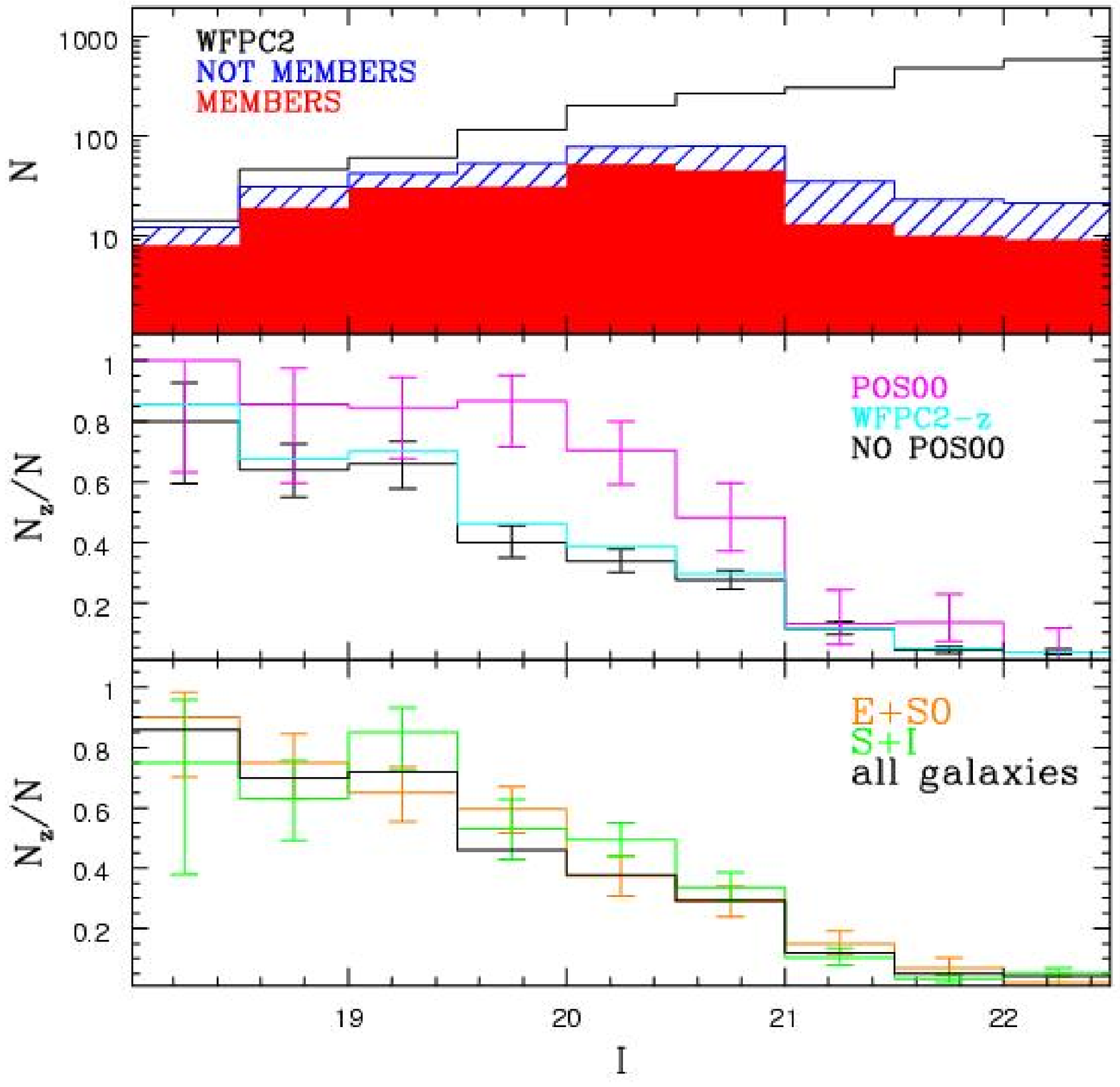}}
\end{center}
\figcaption{Completeness of the WFPC2-z catalog, as a function of $I$
magnitude. Representative error bars are shown for reference. {\bf
Upper panel:} total number of galaxies, number of members
($0.374<z<=0.402$), and number of non-members. {\bf Middle panel:}
redshift completeness, i.e. the fraction of galaxies with measured
redshift, for the entire survey (WFPC2-z) and separately for the central
POS00 field and the remainder of the survey (NO POS00). {\bf Lower
panel:} redshift completeness per broad morphological
class. \label{fig:comp}}
\end{inlinefigure}

\subsection{Redshift Completeness}
\label{ssec:speccomp}

As the redshift catalog involves heterogeneous datasets, it is
important to check redshift completeness in apparent magnitude,
cluster radius and overall spatial extent.

First we examine completeness defined as the ratio of galaxies in the
WFPC2-z catalog (N$_z$) to those in the $I<$22.5 limited WFPC2
catalog, as a function of magnitude and cluster radius. The magnitude
dependence is shown in the middle panel of Fig.~\ref{fig:comp}. This
is very high at bright magnitudes (especially for the central POS00
field), stays above 50 \% until $I\sim21$ and drops significantly
beyond $I\sim21$. The upper panel of the same figure shows the actual
numbers per magnitude bin. Completeness does not depend significantly
on the morphological type, as demonstrated in the lower panel of
Fig.\ref{fig:comp}. Accordingly, in the following analysis we assume
that the WFPC2-z catalog is representative of the entire population
without introducing significant bias in the mix of morphological
types.

\begin{inlinefigure}
\begin{center}
\resizebox{\textwidth}{!}{\includegraphics{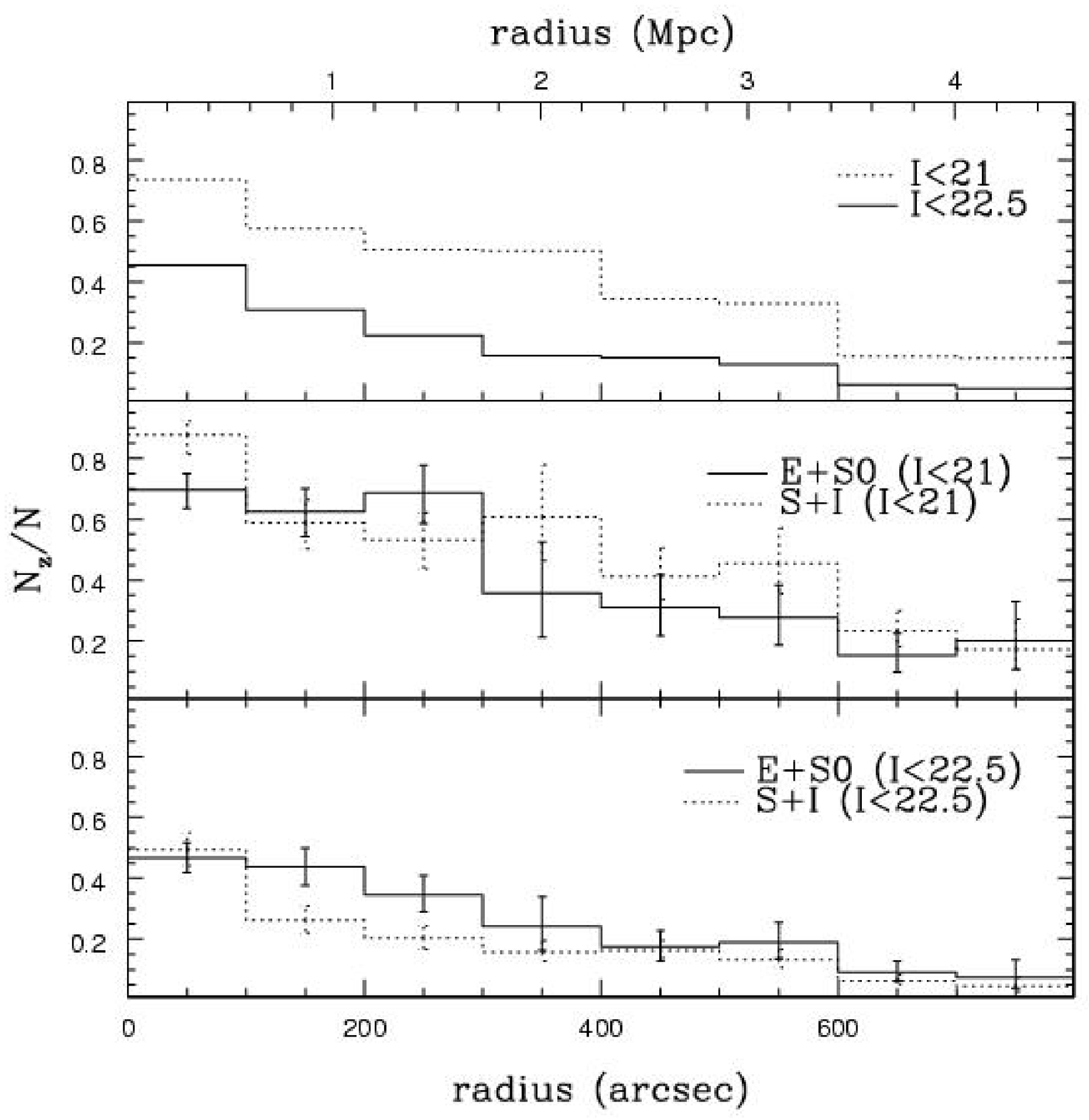}}
\end{center}
\figcaption{Redshift completeness of the WFPC2-z catalog as a function
of cluster radius. On the top axis the scale in Mpc is shown for
reference. Representative error bars are shown for reference. {\bf
Upper panel:} all morphological types.  {\bf Middle panel:} Per broad
morphological type to $I<21$. {\bf Lower panel:} Per broad
morphological type to $I<22.5$.
\label{fig:comp3}}
\end{inlinefigure}

Second, in Figure~\ref{fig:comp3} we show the redshift completeness as
function of cluster radius. At $I<21$ the completeness drops sharply
from almost 80 \% at the center to below 20~\% beyond $600''$ ($\sim$
3 Mpc). At fainter magnitudes the completeness is significantly
smaller, dropping below 20~\% already at $300''$. Completeness as a
function of radius segregated by broad morphological type is shown in
the middle and lower panels of Figure~\ref{fig:comp3}. Importantly, no
major difference is found at either faint or bright magnitude limits,
confirming that the WFPC2-z catalog appears to be representative of
the morphological mix at any given radius.

In summary, the WFPC2-z catalog is sufficiently complete at
$I<21$, particularly in the central Mpc where it is above 60\%. It
declines to 20 \% at 5 Mpc. The completeness is much lower for
galaxies fainter than $I=21$; it is $\sim$ 30 \% for $I<22.5$ in
the central Mpc. Such incompleteness is inevitable even after
considerable efforts. Fortunately, however, the catalog is not
obviously biased in favor of any particular morphological type.

We will argue in the next sections that, by restricting the analyses
to $I\sim21$, uncertainties arising from both incompleteness and
morphological typing (see $\S$ \ref{ssec:unmorph}) are minimal.  We
will assume that the distribution of morphological types in the
WFPC2-z sample is representative. It is recognized that at large radii
(4-5 Mpc) -- where spectroscopic information is most needed to
establish membership -- the WFPC2-z sample is only $\simeq20$~\% of
the parent sample, so significant extrapolation is necessary.
Accordingly, results based on the WFPC2-z catalog in this region must
be interpreted with greater caution.

\section{The Physical Nature of Environmental Evolution in Cluster Galaxies}

\label{sec:mechanisms}

Before interpreting the observations, it is convenient to build a
physical framework for our study by defining the relevant time,
velocity, and length scales associated with some of the various
processes that have been proposed for the environmental evolution of
cluster galaxies. We introduce these processes in
Section~\ref{ssec:zoo}. Since there exists a considerable literature
on this subject with creative terminology often used with various
meanings, we also use this opportunity to define our own
terminology. Ideally we are interested in reconstructing the evolution
of a single, hypothetical representative galaxy as it enters the
cluster potential and interacts with the intercluster medium. To this
aim, we present estimates of the sphere of influence of each relevant
physical process affecting the infalling galaxy.

Using a simple cluster model and estimates of relevant length, time
and velocity scales discussed in the Appendices, in
Section~\ref{ssec:2D} we discuss projection effects and use the
various scales to define various observable regions of the cluster for
each of the physical mechanisms listed in Section~\ref{ssec:zoo}. The
key output is a series of distinguishing zones which will form the
basis of the later analysis.

\subsection{Environmental Processes in Rich Clusters}

\label{ssec:zoo}

For simplicity, we define the various processes under three broad
headings: galaxy-ICM interactions (which refers to the interaction
of a cluster galaxy with the gaseous component of the cluster),
galaxy-cluster gravitational interaction (which includes tidal and
related dynamical processes), and smaller-scale galaxy-galaxy
interactions. Clearly the distinction is, in places, somewhat
arbitrary.

\begin{enumerate}

\item Galaxy-ICM interactions. a) Ram pressure stripping, the removal
of galactic gas by pressure exerted by the intercluster medium (Gunn
\& Gott 1972; Fuijta 1998; Fujita \& Nagashima 1999; Abadi et al.\
1999; Toniazzo \& Schindler 2001; Fujita 2001), serves to terminate
star formation by removing the gas supply. b) Thermal evaporation of
the galactic interstellar medium (ISM) by the hot ICM (Cowie \&
Songaila 1977). c) Turbulent and viscous stripping of the ISM (Nulsen
1982; Toniazzo \& Schindler 2001). d) Pressure-triggered star
formation in which galactic gas clouds are compressed by the ICM
pressure thereby temporarily increasing the star formation rate
(Dressler \& Gunn 1983; Evrard 1991; Fujita 1998).

\item Galaxy-cluster gravitational interactions. a) Tidal compression
of galactic gas (Byrd \& Valtonen 1990; Henriksen \& Byrd 1996; Fujita
1998) by interaction with the cluster potential can increase the star
formation rate; b) Tidal truncation of the outer galactic regions
(e.g. the dark matter halos) by the cluster potential (Merritt 1983,
1984; Ghigna et al.\ 1998; Natarajan et al.\ 1998). If tidal
interactions remove the gas reservoir they can also lead to quenching
of star formation but, more generally, such processes can be inferred
via structural changes in the galactic mass profiles.

\item Galaxy-galaxy interactions. a) Mergers, i.e. low speed
interactions between galaxies of similar mass (Icke 1985; Mihos 1995;
Bekki 1998). b) Harassment, i.e.  high speed interactions between
galaxies in the potential of the cluster (Moore et al. 1996; Moore,
Lake \& Katz 1998; Moore et al.\ 1999).

\end{enumerate}

Depending on the fraction of gas removed and its rate, ram-pressure
stripping, thermal evaporation, turbulent and viscous stripping can
lead either to a rapid quenching of star formation or a slow decrease
in the star formation rate if only the loosely bound reservoir
surrounding field galaxies is affected (Larson, Tinsley \& Caldwell
1980, Balogh, Navarro \& Morris 2000; Diaferio et al.\ 2001; Drake et
al. 2000). Independent of the precise physical process, we will label
this slow decrease in the star formation rate {\it starvation}.

\begin{inlinefigure}
\begin{center}
\resizebox{\textwidth}{!}{\includegraphics{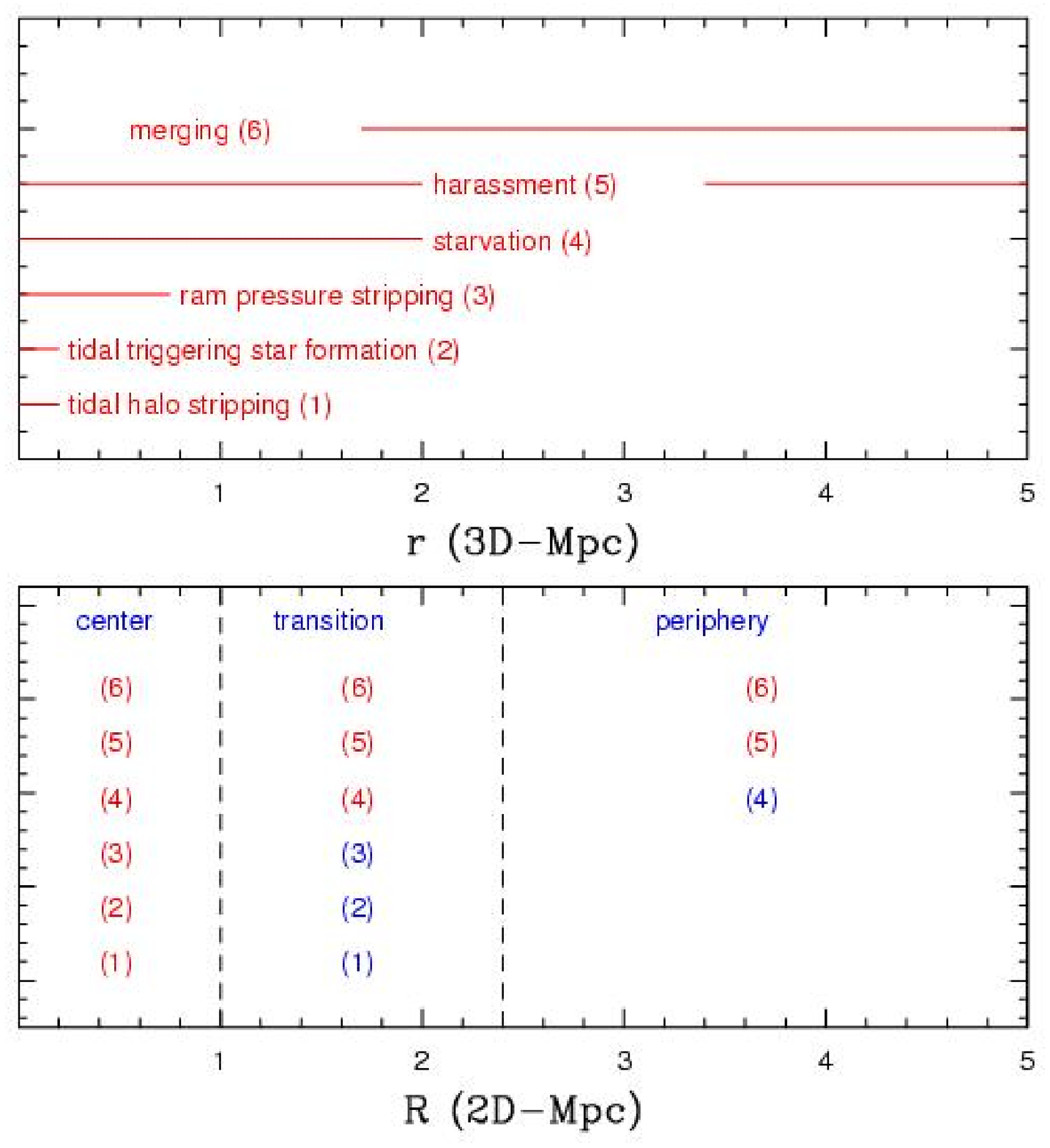}}
\end{center}
\figcaption{Summary of the regions where key physical mechanisms are
likely to operate. Top panel: the horizontal lines indicate the radial
region where the mechanisms are most effective (in 3-D space; note
that harassment is effective in the entire range).  Lower panel:
for each projected annulus (as in Figure~\ref{fig:defrad}) we identify
the mechanisms that can have affected the galaxy in the region
(red). The blue numbers indicate processes that are marginally at
work. For reference, the virial radius is $r_{\rm V}\sim1.7$ Mpc. See
Section~\ref{ssec:TR} for discussion. \label{fig:mechanisms}}
\end{inlinefigure}

The above mechanisms can affect the observed internal structure of
cluster galaxies, including the mass distribution as probed by
galaxy-galaxy lensing, internal kinematics as probed by resolved
spectroscopy, and/or the star formation properties as probed with
diagnostic spectroscopic and photometric features. Crucially, changes
in internal structure and star formation properties can affect
morphologies. In disentangling the various effects, we must attempt to
rank -- at least approximately -- the importance of each mechanism for
various physical regions of the cluster. The dominant physical
mechanisms together with the estimate of their spheres of influence --
as derived in Appendix~\ref{app:ls} -- are summarized in the upper
panel of Figure~\ref{fig:mechanisms}.

\subsection{Defining Diagnostic Cluster Regions}

\label{ssec:2D}

In order to explain the run of observables with radius in terms of the
various physical mechanisms discussed, we use the spheres of influence
shown in Figure~\ref{fig:mechanisms} and the physical scales estimated
in the Appendices, to determine how our instantaneous view of the
observed position of the cluster galaxies may be linked to the
physical and temporal scales of influence of each mechanism taking
into account the effects of projection. Although galaxies can move
significantly during transient phenomena of $\sim 1$ Gyr duration, the
goal is to identify projected regions where we can {\it exclude} the
possibility that certain mechanisms are at work. We define three
annular projected regions: the cluster center, a transition region and
the cluster periphery (shown in Figure~\ref{fig:mechanisms}, lower
panel) within which we will search for correspondence in the observed
properties of the galaxy population.

\begin{enumerate}

\item The central region comprises that within a $\sim 1$ Mpc radius.
It is the region where the cluster potential is steepest and the ICM
is detectable. Most of the mechanisms are effectively at work (see
Figure~\ref{fig:mechanisms}, lower panel), speeds can be high (up to a
few thousands of \kms ), and projection effects can pollute this
central region with galaxies that are at the periphery of the
cluster. Hence unfortunately none of the mechanisms can be definitely
excluded, although we can expect the mechanisms mostly related to the
cluster potential (tidal stripping and tidal triggering) to be
dominant. This is also the region mostly probed by previous HST
surveys.

\item The transition region - an annulus between $\sim 1$ and $2.4$
Mpc comprising the virial radius -- is sufficiently far from the
center that galaxies observed in this region cannot have experienced
tidal effects more recently than 0.5-1 Gyr ago and cannot be
experiencing ram-pressure stripping (although they can have in the
recent past).

\item The periphery is the annulus between $\sim2.4$ Mpc and the
outermost radius probed ($\sim$ 5 Mpc). It is so far from the center
(and speeds are lower than in the rest of the cluster) so that, given
the discussion in Appendix~\ref{app:tvs}, we can safely assume that
most of the galaxies at the periphery of the cluster {\it have never
been through the cluster center} and therefore have never experienced
the effects of tidal stripping, tidal triggering of star formation and
ram-pressure stripping.  Nevertheless, they may be experiencing
starvation and are able to undergo mergers or be harassed by
massive nearby galaxies.

\end{enumerate}

\section{The Morphological Distribution in Cl0024}

\label{sec:morph}

We now combine the morphological and the redshift catalogs to
study the distribution of morphological types in Cl0024+16 both as
a function of radius (hereafter the T-R relation;
Section~\ref{ssec:TR}) and as a function of the local projected
density (hereafter the T-$\Sigma$ relation; Section~\ref{ssec:TS})

In the local Universe, the T$-\Sigma$ and T-R relations have been
actively studied over the past three decades (Oemler 1974; Melnick \&
Sargent 1977; Dressler 1980; Postman \& Geller 1984; Giovanelli et
al.\ 1986; Whitmore \& Gilmore 1991; Oemler 1992; Whitmore, Gilmore \&
Jones 1993), often in terms of the quest for the most fundamental
relation. Indeed, there is a strong scientific motivation for
determining and comparing these relations over a wide range of
environmental density in order to determine which physical processes
fundamentally governs the morphological evolution of galaxies.  For
example, the T-R relation should provide a tool to investigate the
effects of those phenomena which are related to the cluster's
gravitational potential including interaction with the hot
intercluster medium: such phenomena should run broadly as a function
of azimuthally-smoothed radius. On the other hand, the effects of
local overdensities and subclustering in the resident or newcomer
population, will be erased in the T-R analysis. In this case, it is
useful to examine the T-$\Sigma$ relation.

Dressler (1980) introduced the T-$\Sigma$ relation as a means to
investigate the environmental effects on galaxy evolution in non
symmetric (``non-relaxed'') clusters, where the definition of a cluster
center was difficult and ambiguous. Remarkably, Dressler (1980) found
that the T-$\Sigma$ relation in the local Universe was identical
for ``relaxed'' and ``non-relaxed'' clusters (see also Dressler et
al.\ 1997), and used this result to argue against ram-pressure
stripping as a dominant mechanism to transform spirals into
E+S0. Whitmore \& Gilmore (1991) and Whitmore et al.\ (1993)
reanalyzed the data in the Dressler's sample applying several
corrections to the background estimate, magnitude limit, and center
estimate.  They claimed that morphology correlates better with radius
than with local density and since a sharp decrease in the fraction of
E+S0 with radius is seen only in the innermost regions ($\sim 0.5$
Mpc) they argued that tidal disruption of spirals and S0 by the
cluster potential is the dominant mechanism.

Further insight can be gained by studying the T-R and T-$\Sigma$
relation at intermediate redshift $z\sim0.5$, where the galaxy
population is undergoing profound transformations. Dressler et al.\
(1997) studied the T-$\Sigma$ and T-R relations in the core (typically
within the inner 0.5 Mpc) of a sample of clusters at $z\sim0.3-0.5$
including Cl0024+16. For the high concentration clusters -- such as
CL0024+16 -- they found the fraction of early-type galaxies was a
steeply increasing (decreasing) function of local density (radius). By
contrast, the low concentration clusters did not show any gradient in
the fraction of early-type galaxies with respect to local density or
radius. Consequently they inferred that low concentration clusters are
less evolved then high concentration ones and segregation of
morphological types has not happened yet. Unfortunately, their data
did not cover the periphery of clusters, which we argue
(Section~\ref{sec:mechanisms}) is helpful in understanding the
mechanisms driving the change of the morphological mix between
$z\sim0.5$ and today (Dressler et al.\ 1994; Couch et al.\ 1994;
Dressler et al.\ 1997).

\begin{inlinefigure}
\begin{center}
\resizebox{\textwidth}{!}{\includegraphics{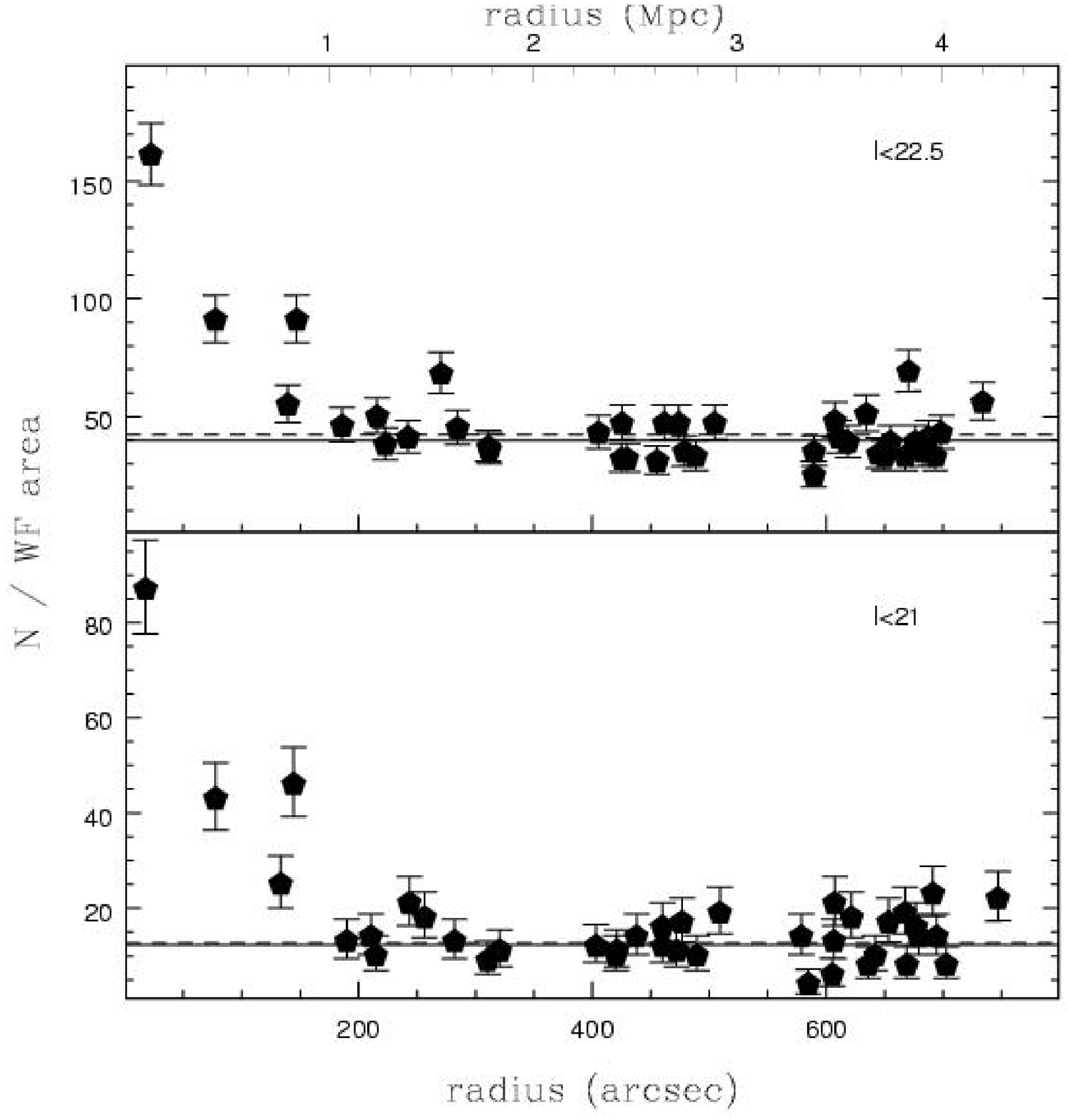}}
\end{center}
\figcaption{Average observed galaxy density per WFPC2 pointing in the
Cl0024+16 field as a function of distance from the cluster center for
two magnitude limits. The top axis shows the scale in Mpc. Error bars
show the errors computed assuming Poisson noise.  Solid and dashed
lines indicate field number counts determined respectively by Abraham
et al.\ (1996a) and Postman et al.\ (1998).  Beyond $\sim 1$ Mpc the
surface density approaches this value.
\label{fig:densR}}
\end{inlinefigure}

We wish to extend the T-$\Sigma$ and T-R relations to larger radii (5
Mpc) and lower projected densities. As a laboratory for investigating
this issue, Cl0024+16 is an interesting case because the two
descriptions are not completely equivalent. Both the redshift
distribution (Figure~\ref{fig:histoz}) and the projected distribution
(Figure~\ref{fig:spacedistAB}) indicate the cluster is not perfectly
regular or relaxed (Czoske et al.\ 2002), notice in particular the
overdensity NW of the center, with a somewhat poor correspondence
between radius and projected density. These issues are shown
quantitatively in Figure~\ref{fig:densR}, where we plot the observed
(not field-subtracted) average galaxian density per WFPC2 pointing as
a function of cluster radius. Although an average trend is seen, there
is considerable scatter (Section~\ref{ssec:TS}). Accordingly,
contrasting galaxy properties in the context of both variables will be
illuminating.

\subsection{Measuring the T-$\Sigma$ and T-R relations}

\label{ssec:backfore}

In order to derive the T-$\Sigma$ and T-R relations we must remove
background/foreground contamination, compute the limiting
magnitude and determine the cluster center. The ideal solution for
resolving field contamination - the spectroscopic identification
of all cluster members to our magnitude limit over the entire
Cl0024+16 field - is beyond our current observational resources.
Despite almost 800 redshifts, our WFPC2-z catalog is only 20 \%
complete at the periphery (Section 3). Achieving reasonable
spectroscopic completeness over such a large area is certainly a
worthwile goal within reach of the new generation of
high-multiplexing spectrographs currently being commissioned on
large telescopes (DEIMOS \& IMACS on Keck and Magellan, for
example). For now, we adopt a statistical approach based on the
average field number counts. However, despite progress in
measuring field counts (e.g. Postman et al.\ 1998; Casertano et
al.\ 2000), the uncertainties are still significant for our
purposes. In Figure~\ref{fig:densR} we plot both the I-band
background surface density as measured by Abraham et al.\ (1996a)
and Postman et al.\ (1998), to illustrate this uncertainty. The
uncertainty is clearly relevant at large radii. Additionally,
field galaxies are clustered on the angular scale of a WF chip
(see also Valotto, Moore \& Lambas 2001) thus field to field
variation dominates the uncertainty in the background
contamination.

To estimate the average background count per morphological type,
we adopt the Postman et al.\ (1998) number counts, scaled by the
fraction of morphological types found by Abraham et al.\ (1996a)
at the corresponding limiting magnitude. For example the fraction
of compact/E+S0/S (see below) are 5\%:28\%:49\% at $I<21$ and
10\%:18\%:43\% at $I<22.5$. This procedure combines the advantage
of having the large area used by Postman et al.\ (1998) to
determine the counts, with the background morphological
distribution determined on the same scheme by the same classifier.

\medskip

Another important issue is the computation of the limiting
absolute magnitude. The catalog of local clusters used by Dressler
et al.\ (1980) is limited to galaxies more luminous than
$M_{V}=-20.4$ ($H_0=50$ \kms Mpc$^{-1}$; see also Dressler et al.\
1997 and references therein). This limit can be converted into an
I-band apparent magnitude using,
\begin{equation}
M_{\rm V}=I-DM+\Delta m_{\rm V8},
\label{eq:V8}
\end{equation}
where the $k$-color correction (Treu et al.\ 2001a) $\Delta
m_{\rm V8}=0.85\pm0.03$ is computed using a broad range of synthethic
spectra (Bruzual \& Charlot 1993; GISSEL96 version) and empirical
templates (Kinney et al.\ 1996). The uncertainty in the transformation
is very small because, at the redshift of Cl0024+16, the F814W filter
closely matches rest frame $V$. From Eq.~\ref{eq:V8} the apparent
magnitude limit of the local morphology-density analysis by Dressler
et al.\ (1980) corresponds to I=21.13.

Since a galaxy with $M_{\rm V}=-20.4$ at $z$=0 most likely did not
have the same luminosity at $z\sim0.4$, it is important to consider
corrections for luminosity evolution. For the E+S0 population,
Fundamental Plane (Djorgovski et al.\ 1987; Dressler et al.\ 1987)
studies indicate rest-frame $V$ band brightening of $\sim0.4-0.6$ mags
by $z$=0.4 (van Dokkum \& Franx 1996, Treu et al.\ 1999, 2001b; van
Dokkum et al.\ 2001; Bernardi et al.\ 2003). The expected evolution
for spirals is less clear, with reported measurements ranging from
none to up to $\sim1$ magnitude of brightening to $z\sim1$ in
rest-frame $B$ (Lilly et al.\ 1998; Simard et al.\ 1999). Metevier,
Koo \& Simard (2002) have constructed the Tully-Fisher relation for 7
galaxies in Cl0024+16 and find no evidence for a change in the slope
or intercept with respect to local samples. Clearly if these
indications are correct and we ignored any luminosity correction, we
would overestimate the fraction of early-type galaxies. Given the
uncertainties, as in previous studies we prefer to present the
observations, neglecting evolution. Fortunately, our results are not
particularly sensitive to changes of $\sim 0.5$ mags in the applied
limit (changes within the errors; see also Dressler et al.\ 1997;
Kodama \& Smail 2001). Furthermore, we are particularly interested in
the trends within one system, which are not affected by differential
luminosity evolution.

\medskip

Although Cl0024+16 is centrally concentrated and approximately
rotationally symmetric to the extent that a T-R analysis is
justified, some asymmetry is seen, e.g. the overdensity to the NW.
For consistency with previous studies we adopt the X-ray center as
cluster center. However, we have repeated the analysis in the
following sections adopting alternative centers (Table 1) and the
results are robust, changing less than the error bars.

\medskip

Finally, we will restrict ourselves to the broad morphological
classification scheme which is robust at $I<21.1$
(Section~\ref{ssec:WFPC2morph}). This provides good statistics at
low densities/large radii and enables us to perform better field
subtraction. For simplicity we will refer to the broad
morphological classes as compacts (C;T=-1), early-types
(E+S0; 0-2), spirals (S; 3-5), and ``irregulars'' (I; 6).  Finer
classification, including the separation of E and S0 galaxies
(see, e.g., Dressler et al.\ 1997, Andreon et al.\ 1998; Fabricant
et al.\ 2000) will be deferred until a later paper.

\subsection{Radial trends}

\label{ssec:TR}

\subsubsection{Observations}

We now discuss the observed radial trends in morphological types.
There are three key issues. How far out in radius can we trace the
cluster in our dataset? What is the run with radius of the density
of E+S0 and S+I galaxies? How does the cluster population of
Cl0024+16 compare to the surrounding field?

The lower panel of Figure~\ref{fig:defrad} shows the fraction of
spectroscopic members (defined as that fraction of all with redshifts
lying within $0.374<z<0.402$) as a function of radius.  Error bars
were obtained assuming the binomial distribution (Gehrels
1986). Clearly, there is an excess of galaxies to the largest radius
($\simeq$5 Mpc). Using the Canada France Redshift Survey Spectroscopic
catalog (Lilly et al.\ 1995) limited at $I<21.1$, only a few percent
of field galaxies is expected to lie within this narrow redshift
range. In the two upper panels we show the E+S0 and S+I surface
density as a function of radius, computed by removing the background
as discussed above (Sec~\ref{ssec:backfore}). Error bars are computed
assuming Poisson statistics (Gehrels 1986) not corrected for field
galaxy clustering. At large radii (the periphery) no
statistically-significant excess of either type is found. This is due
in part to the large additional uncertainties involved in the
type-dependent subtraction. A small systematic over-subtraction of the
background cannot be excluded, considering the cluster is
optically-selected and the presence of a local background underdensity
would have enhanced its discovery (Dressler 1984).  The type-dependent
excess is detectable out to a radius of $1-2$ Mpc and becomes very
significant only within the inner 1 Mpc. As far as the mix of
morphological types is concerned, the densities of E+S0 and S+I galaxies
remain quite similar down to the central regions and only within the
central 200 kpc does the E+S0 component become the dominant component.

\begin{inlinefigure}
\begin{center}
\resizebox{\textwidth}{!}{\includegraphics{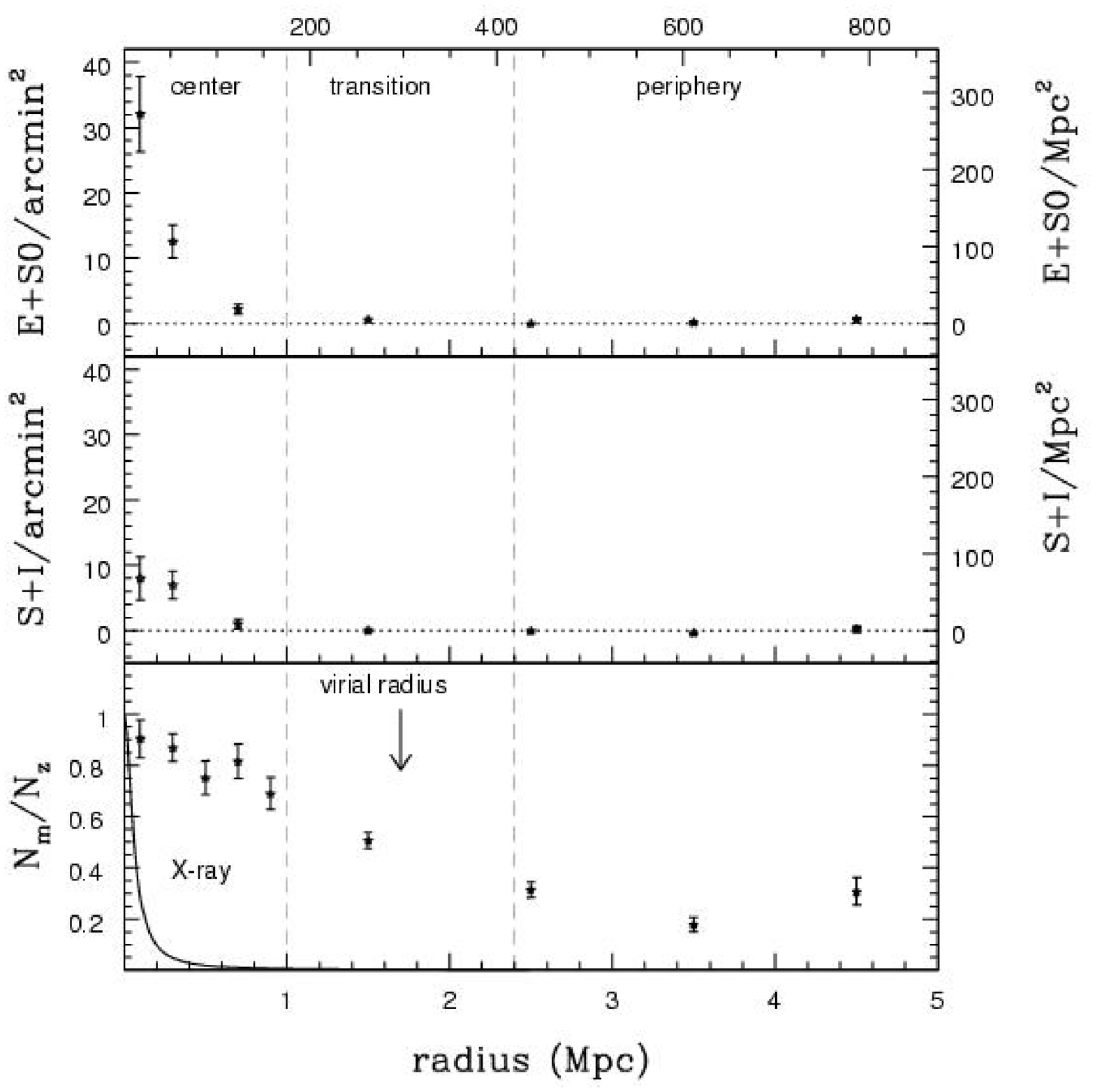}}
\end{center}
\figcaption{Radial trends. {\bf Top panel:} Surface density of member
E+S0 galaxies ($I<21.1$) as a function of radius, after statistical
removal of the background/foreground contamination. The density in
galaxies per Mpc$^2$ is shown on the right scale for reference.  Error
bars assume Poisson statistics. {\bf Middle panel:} as the top panel
for spiral and irregulars members. {\bf Lower panel:} fraction of
galaxies with redshifts (N$_z$) that are members (N$_m$) as a function
of radius. Note that members are found to the outermost region
probed. The three diagnostic regions defined in
Section~\ref{sec:mechanisms} and Figure~\ref{fig:mechanisms} are
indicated. The best fitting X-ray surface-brightness $\beta$ profile
(B\"ohringer et al.\ 2001), in arbitrary units, is shown on the lower
panels for comparison. \label{fig:defrad}}
\end{inlinefigure}

Figure~\ref{fig:defrad} shows the importance of using
spectroscopically confirmed members for the periphery. In the inner
Mpc however, the error introduced by our statistical
background/foreground correction is acceptable and both the WFPC2 and
the WFPC2-z catalogs can be used. The result of enlarging the inner
sample to include the WFPC2 catalog is shown in
Figure~\ref{fig:earlyfrac} for both E+S0 (upper panel) and S+I (lower
panel) galaxies. Fractions determined from the WFPC2-z sample are
plotted as solid symbols, with errors determined assuming a binomial
distribution. The fractions could be biased if the incompleteness
depends on morphological types, which seems not to be the case
(Sec~\ref{ssec:speccomp}). However, the numbers involved are small
($\sim$ 50 galaxies per radial bin) and the errors accordingly
large. Fractions determined from the WFPC2 sample (open points) suffer
less from small number statistics, therefore are probably to be
preferred in the innermost regions ($\la 0.5$ Mpc) when the
uncertainty due to background/foreground contamination removal is
sufficiently small.

The inner Mpc shows a steep gradient in morphological mix. E+S0
galaxies dominate and reach $\sim 75$\% of the population in the
inner 200 kpc (lower than in local clusters where they reach
almost 100\%, e.~g. Whitmore, Gilmore \& Jones 1993). Beyond 1 Mpc
the fraction of E+S0 remains constant within the error, or at most
mildly declining (from $51\pm7$\% to $42\pm13$\%). A more
significant difference is seen between the morphological mix at 1
Mpc ($51\pm7$\%) and the field ($37\pm7$\%). The field fraction
was determined by combining the subsample of the CFRS-LDSS with
HST images (Brinchmann et al.\ 1998), limited to $I<21.1$ and to
$0.3<z<0.5$, interlopers located in the Cl0024+16 WFPC2-z catalog
with $I<21.1$, $0.3<z<0.37$ or $0.42<z<0.50$, and interlopers
located in the Morphs database with HST images in F814W and
$I<21.1$, in the range $0.3<z<0.5$. These field samples are the
most appropriate given that they were consistently classified by
the same classifier (RSE), and are either spectroscopically
complete (CFRS-LDSS) or at least suffer from similar
incompleteness as the cluster samples (Cl0024+16 and the Morphs).
However, the mix in the field sample is affected by small number
statistics\footnote{The numbers are E+S0/S+I/Total 24/39/64 i.e.
$37\pm7$\%,$61\pm7$\%, assuming binomial distribution.} and cosmic
variance. Surprisingly, no larger spectroscopically complete
sample of field galaxies with HST imaging is yet available to
improve this comparison. The accuracy of the cluster fractions
will improve when more redshifts in the periphery will be
available, allowing us to improve the determination of the slope
of the early-type fraction from the transition region to the
periphery to the field.

\begin{inlinefigure}
\begin{center}
\resizebox{\textwidth}{!}{\includegraphics{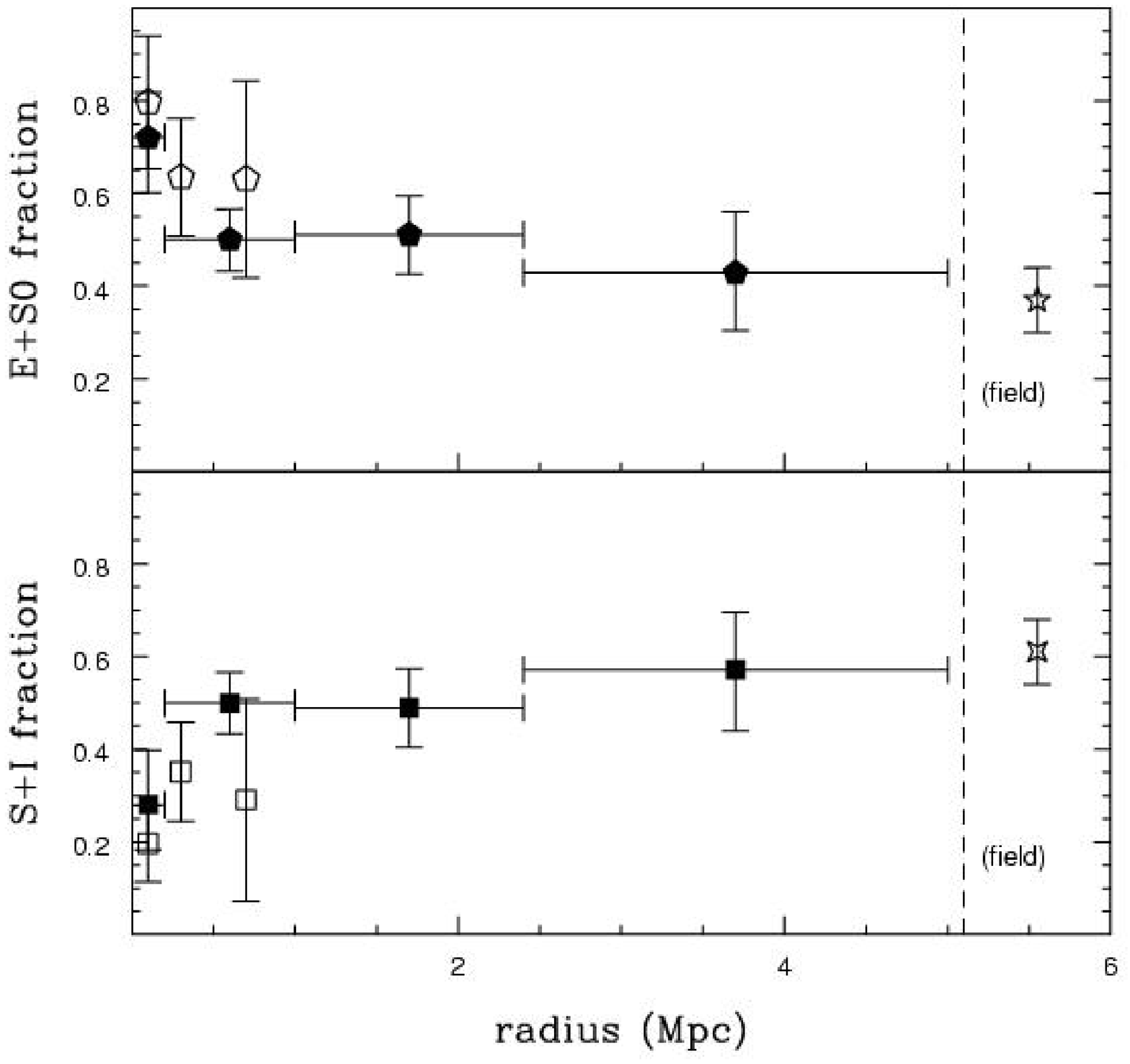}}
\end{center}
\figcaption{The fraction of morphological types ($I<21.1$) in
Cl0024+16 as a function of cluster radius: (upper panel) fraction of
E+S0 galaxies; (lower panels) fraction of spirals. Fractions
determined from the entire WFPC2 sample -- removing the background
statistically as detailed in Section~\ref{ssec:backfore} -- are shown
as large empty points, while fractions determined from the
spectroscopic WFPC2-z catalog are shown as solid points. Points beyond
the dashed line at 5.1 Mpc are field fractions determined from the
CFRS-LDSS survey ($0.3<z<0.5$ ; Brinchmann et al.\ 1998), field
galaxies in the Cl0024+16 area in the cluster vicinity ($0.3<z<0.37$
or $0.42<z<0.4$), and interlopers with F814W imaging from the Morphs
collaboration (Smail et al.\ 1997; Dressler et al.\ 1999).
\label{fig:earlyfrac}}
\end{inlinefigure}

\subsubsection{Discussion}

\label{sssec:TRdisc}

As a first step in understanding the radial run of the E+S0
fraction, we seek to identify which physical mechanism(s) among
the ones described in Section~\ref{sec:mechanisms} can (or cannot)
be responsible for the transformation of the infalling galaxies.
To investigate this, we will consider the three diagnostic regions
defined in Section 4 (Figure~\ref{fig:mechanisms} and
Figure~\ref{fig:defrad}): the center, transition, and periphery.

In the transition region, isolated infalling galaxies cannot have
experienced tidal effects or ram pressure stripping more recently than
0.5-1 Gyr ago. In the periphery most of the galaxies {\it have never
been through the center} and therefore are free from the effects of
tidal stripping, tidal triggering of star formation and ram-pressure
stripping. The mild decline in star formation associated with the
gentle demise of spirals between the field and the transition region
(Figure~\ref{fig:earlyfrac}) cannot be ascribed to tidal interactions
with the cluster potential nor to ram pressure stripping or triggering
of star formation none of which are effective here. Setting aside
merging -- which we will discuss later -- this leaves us starvation
and harassment. Ultimately, a direct observation that the infalling
population has smaller gas reservoirs than in the field would be
needed to discriminate these. Accurate star formation diagnostics at
large radii would provide some constraints. If significant numbers of
starbursting/post-starburst galaxies (see Dressler et al.\ 1999;
Poggianti et al.\ 1999) are found at the periphery, quenching would
have to occur after a burst had depleted the gas, in contrast with the
starvation picture (Balogh, Navarro, Morris 2001; Diaferio et al.\
2001). Harassment should be detectable through an increased fraction
of disturbed/interacting galaxies with respect to the field (Oemler et
al.\ 1997). Unfortunately, it is not obvious how to unambiguously
recognize the effects especially for high surface brightness Milky-Way
type spirals where the signatures could be subtle (Moore et al.\
1999).

Whichever mechanisms are involved, the timescales for the gradual
morphological transformation between the field and the transition
region are rather slow (see Section~\ref{sec:mechanisms} and
Appendix~\ref{app:tvs}). Are the same phenomena, appropriately
accelerated and intensified, responsible also for the steep trend
in the inner Mpc?  Or are additional mechanisms at work? A
quenching mechanism could produce gentle morphological changes
over a few Gyr and the mild gradients observed beyond 1 Mpc
radius. An additional mechanism, more appropriate to the cluster
center, would induce rapid and dramatic changes via bursts of star
formation and/or major structural changes. Multiple
mechanisms/timescales have been discussed by Poggianti et al.\
(1999) and Dressler et al.\ (1999) who claim that the star
formation properties of infalling galaxies react rapidly to
environmental influences becoming visible as soon as they sense
the cluster potential. By contrast, more time is needed for
significant morphological transformations.

A completely different possibility discussed in the next section
involves a segregation effect. Objects at different radii represent
ones with different early assembly (merger) histories which are
insufficiently mixed in the cluster to erase correlations between
their history and their current location (Diaferio et al.\ 2001).

\subsection{The morphology-density relation}
\label{ssec:TS}

\subsubsection{Observations}

We now examine the $T-\Sigma$ relation to further investigate the
infalling scenario and the effects of segregation. Following
Dressler (1980), we estimate the local projected density
associated with each galaxy by computing the area of the rectangle
with sides parallel to the cardinal axis that encloses the galaxy
and its ten nearest neighbors, using this area to compute the
local density, and finally by removing the background
contamination (as described above in Section~\ref{ssec:backfore}).
We only consider those galaxies associated with a positive
density.

\begin{inlinefigure}
\begin{center}
\resizebox{\textwidth}{!}{\includegraphics{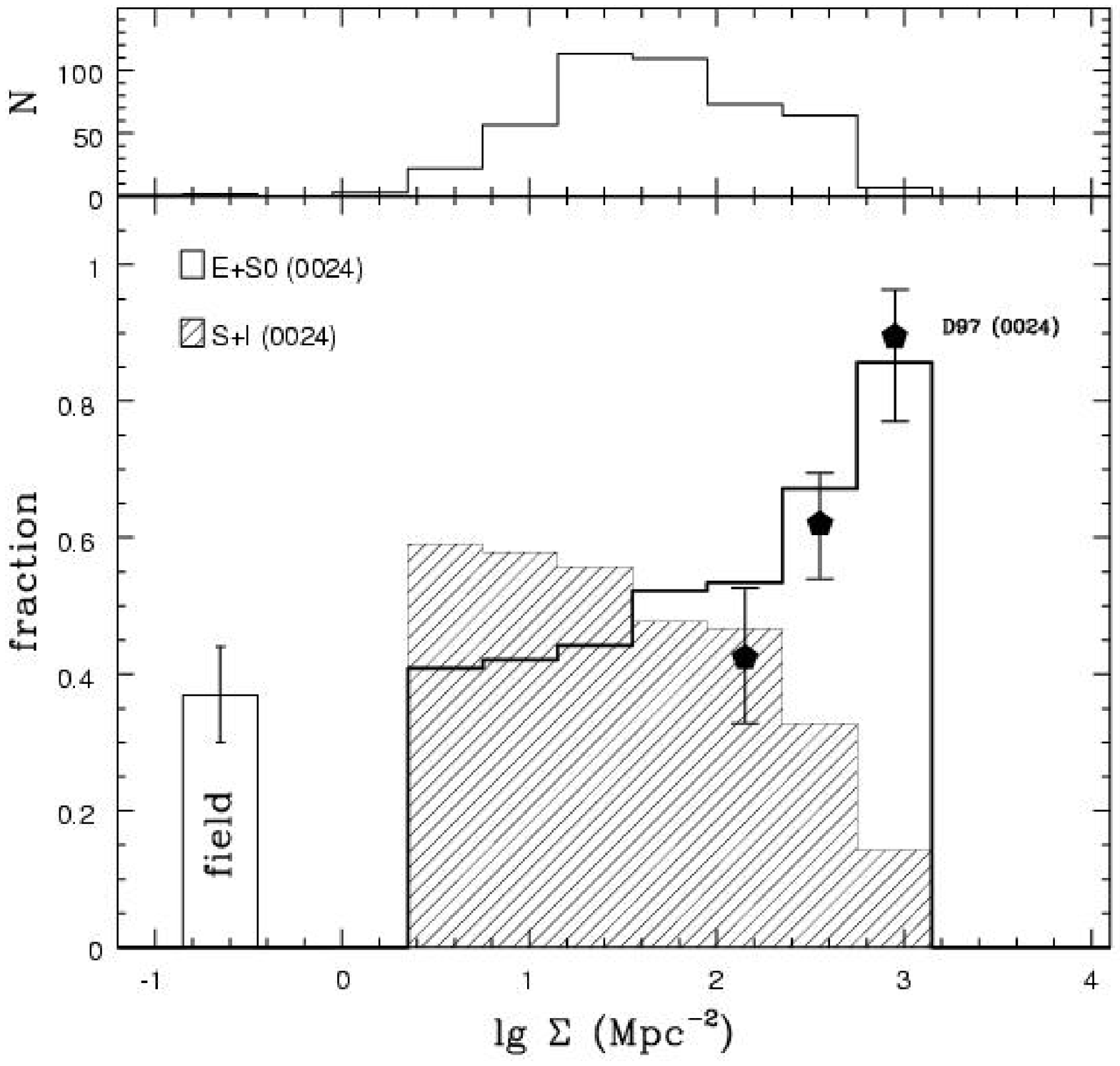}}
\end{center}
\figcaption{Morphology-projected density relation for galaxies in
  Cl0024+16. For comparison, points with error bars show the fraction
  of E+S0 determined by the Morphs collaboration for the central field
  of Cl0024+16 (Smail et al.\ 1997; Dressler et al.\ 1997). The
  histogram with error bar at low surface densities represents the
  fraction of E+S0 in the field at $z\sim0.4$ (at $I<21.1$; see
  Section~\ref{ssec:TR} for discussion).
\label{fig:TScfrM}}
\end{inlinefigure}

The resulting T-$\Sigma$ relation (Figure~\ref{fig:TScfrM}) shows
that the fraction of early-type galaxies increases monotonically
with local surface density. For comparison, we overplot the
fraction of E+S0 as a function of projected density for the
central region of Cl0024+16 as determined by the Morphs
collaboration (Smail et al.\ 1997; Dressler et al.\ 1997)
\footnote{A very similar relation is obtained by combining data
from the three high-concentration clusters observed by the Morphs
collaboration in the F814W filter, i.~e. Cl0024+16, Cl0016+16,
Cl0054-27.}, as appropriate for our assumed cosmology and
magnitude limit.

Our data extends the measurement by well over an order of
magnitude in projected density $\Sigma$. In the region of overlap
there is a good agreement between the two measurements. The slope
in the fraction of E+S0 galaxies appears to flatten just below the
limit probed by the Morphs, declining gently from $\sim 50$\%
towards the field\footnote{The surface density of the field has
been estimated as the counts yielded by the galaxy luminosity
function (e.~g. Ellis et al.\ 1996) integrated 10 Mpc along the
line of sight.} value of $37\pm7\%$ (obtained as described in
\S~\ref{ssec:TR}).  This behavior is very similar to that observed
locally, where the T-$\Sigma$ relation extends over several orders
of magnitude in density from clusters continuously to the field
(Postman \& Geller 1984; see also Bhavsar 1981, de Souza et al.\
1982, Giovanelli et al.\ 1986).

\subsubsection{Discussion}

\label{sssec:TSdisc}

The existence of the T-$\Sigma$ relation at low densities was not
{\em a priori} implied by the existence of the $T-R$ relation,
even for an approximately regular cluster such as Cl0024+16.
Figure~\ref{fig:SvsR} shows that local density is not a well
defined function of radius, except in the central $\sim 0.5$ Mpc.
Spikes in projected density -- such as that at $\sim$ 1 Mpc --
arise as a result of substructures (a similar phenomenon is seen
also locally, see e.g. Andreon 1996).

\begin{inlinefigure}
\begin{center}
\resizebox{\textwidth}{!}{\includegraphics{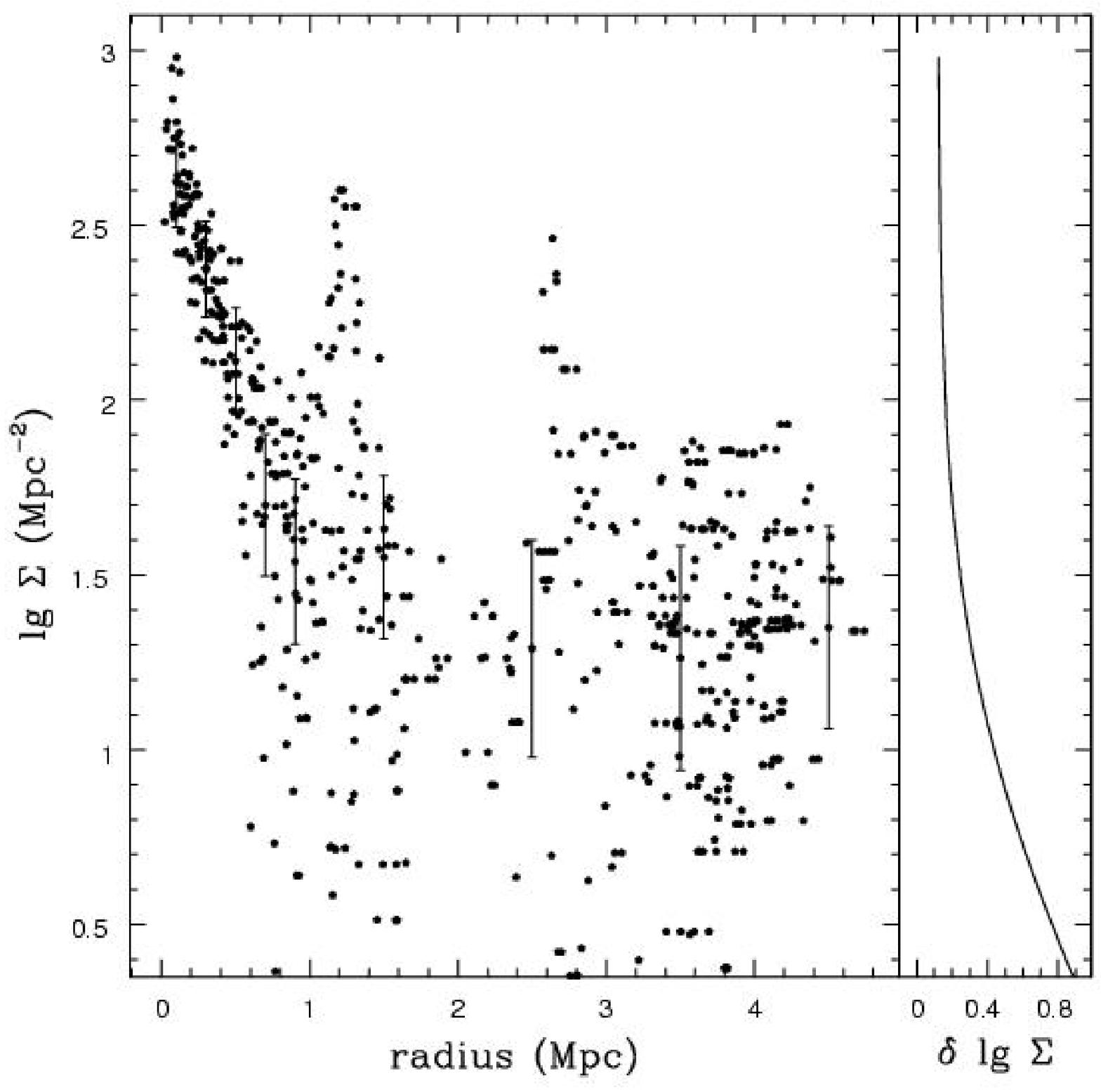}}
\end{center}
\figcaption{Local surface density vs cluster radius in Cl0024+16
revealing the large range at a given radius and the spike in local
density at $\sim 1$ Mpc. The points with error bars indicate the
average $\Sigma$ within radial bins, and the scatter expected from
measurement errors, based on Poisson statistic for the cluster members
and on the uncertainty on the background removal based on the
two-point angular correlation function measured by Postman et al.\
(1998). The expected uncertainty as a function of $\Sigma$ is shown in
the right panel.
\label{fig:SvsR}}
\end{inlinefigure}

The scatter is significantly larger than expected from measurement
errors. Using the definition of $\Sigma$ (Dressler 1980), the
measurement scatter can be estimated assuming a Poisson
distribution for the cluster members and the field to field
variation of the background, corrected using the two-point
correlation function as measured by Postman et al.\ (1998). The
expected error as a function of $\Sigma$ is shown in the right
panel of Figure~\ref{fig:SvsR}, while the points with error bars
represent the average $\Sigma$ in radial annuli, with the scatter
expected from measurement uncertainties.

How is the scatter in the local density at a given radius consistent
with the derived T-$\Sigma$ relation? At face value the existence of a
relation implies galaxies are more aware of their local overdensity
than cluster location. We investigated this further by dividing our
sample in two subsets\footnote{We also tried diving the sample in two,
inside and outside 0.5 Mpc, and the results are very similar to the
case discussed here.}: one from the central POS00 field (i.e the
inner 0.5 Mpc, where the $\Sigma-R$ relation is monotonic) and one
from the overdensity $\sim1$ Mpc NW of the center corresponding to the
spike in local density in Figure~\ref{fig:SvsR} (POS36, see map in
Figure~\ref{fig:map}).

\begin{inlinefigure}
\begin{center}
\resizebox{\textwidth}{!}{\includegraphics{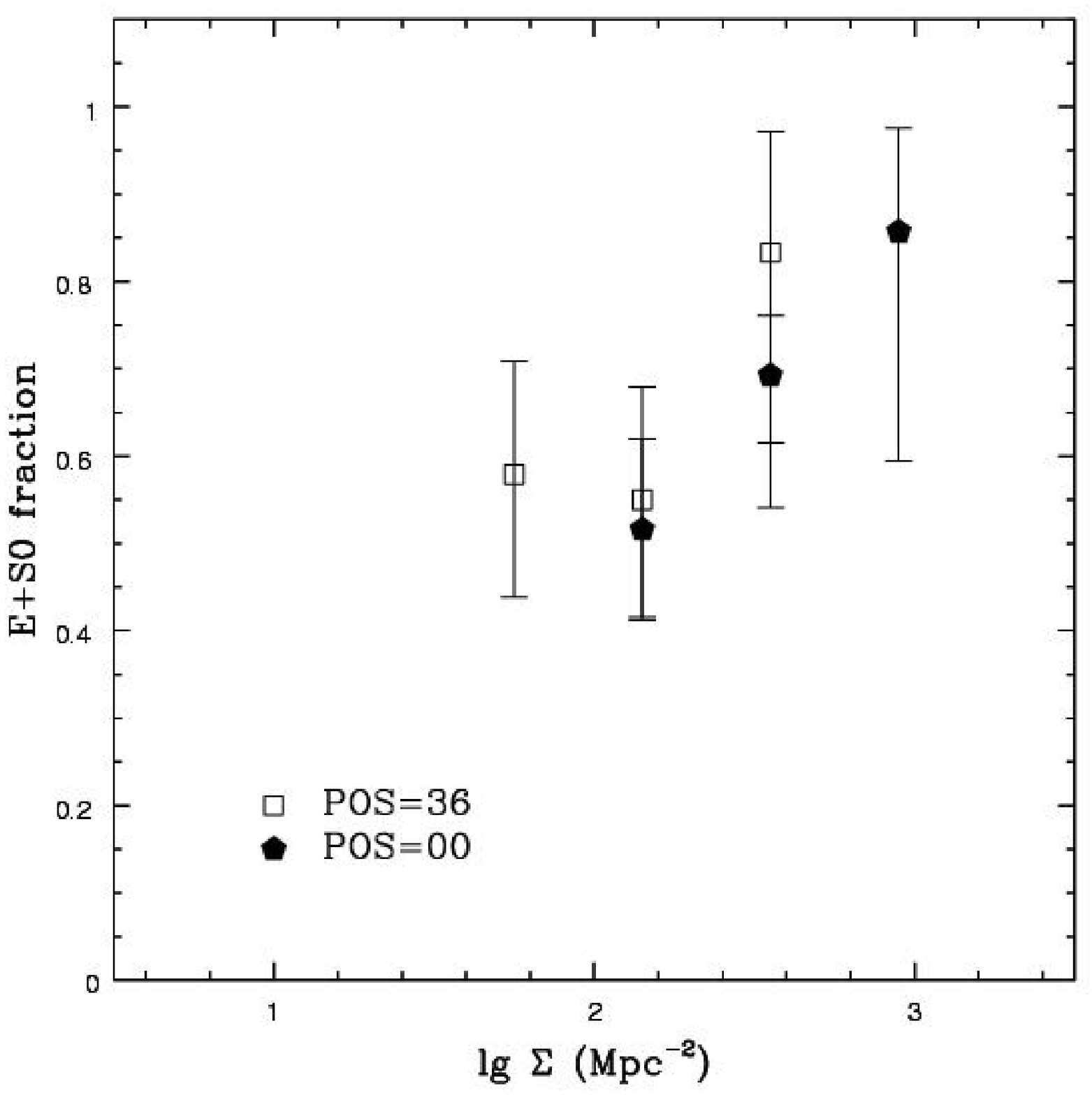}}
\figcaption{The fraction of E+S0 galaxies in Cl0024+16 as a function
of local projected density for galaxies in the central (solid
pentagons) and in the NW overdensity region (open squares; POS36)
corresponding to the spike in local density at $\sim 1$ Mpc in
Figure~\ref{fig:SvsR}.  Note the close match in the regions of
overlapping density.
\label{fig:TSR}}
\end{center}
\end{inlinefigure}

Figure~\ref{fig:TSR} shows the results of this exercise. The
fraction of early-type galaxies still correlates with the local
density in the two subsets, i.~e. galaxies at large radii clearly
``know" if they are in a local overdensity. In the overlapping
density region, the fractions of morphological types are mutually
consistent. These conclusions are particularly significant when
one considers that the cluster galaxies within $\sim 1$ Mpc are
moving at speeds of $\sim 1$ Mpc Gyr$^{-1}$. At the typical
densities of $\sim 100$ gal Mpc$^{-2}$ where differences in the
morphological mix become significant, the smoothing box inherent
to our measurement is of order $\sim 0.1$ Mpc$^{2}$. Hence
galaxies with uncorrelated velocity vectors would erase the
T-$\Sigma$ relation in much less than a Gyr, unless we assume that
morphology is a transient phenomenon with implausibly short
duration.

Not only are galaxy properties arranged according to their local
environment but apparently in undertaking their motion within the
cluster potential they {\it move with it} (i.e. they retain their
identity in these groups as infall continues). This is a clear
indication that -- at least outside the central region -- the
cluster is more logically viewed as a collection of larger
self-contained poorly-mixed clumps (Dressler \& Shectman 1988,
Zabludoff \& Franx 1993). The presence of substructure is
confirmed at high significance by the Dressler \& Shectman (1988)
test: even considering galaxies in peak A alone the probability of
positions and radial velocities being uncorrelated is less than
$4\times10^{-4}$ ($<10^{-4}$ combining peaks A and B). Another way
to view this result is that, within the error bars, outside the
cluster center, the mix of morphological types at given surface
density is irrespective of that clump's location within the
cluster (see Figure~\ref{fig:TSR}).

Although radial trends in morphology were discussed in Section 6.2
in the context of infall timescales, we have argued on the basis
of Figures 14-16, that outside 1 Mpc, the local projected density
plays the main role.  We interpret this as evidence that Cl0024+16
is accreting galaxies in organized substructures, each of which
obeys some form of T-$\Sigma$ relation. An interesting explanation
of our observations would be that the correlation between local
density and morphology arises as a result of pre-cluster
conditions, possibly because galaxies located at different local
densities suffered different assembly and star formation
histories.

Both radial and density effects are important but influence different
regimes. Beyond $\sim0.5-1$ Mpc, we are witnessing trends that relate
to the previous assembly histories of infalling groups. Starvation may
play some role in changing the fate of galaxies in these clumps, but
whatever process is at work, it operates slowly and is insufficient to
supplant that based on segregation which is not erased by
mixing. Inwards of $0.5-1$ Mpc, the tight correlation between radius
and density (Figure~\ref{fig:SvsR}) presumably follows the disruption
of this substructure. At this level, the correlation with the initial
local environment is probably lost, while retaining the morphological
segregation as a function of radius. The cluster radius then becomes
the fundamental variable. Phenomena more closely related to the
cluster potential -- such as tidal interaction or ram pressure
stripping -- may play an additional role in morphological
transformations.

The $T-\Sigma$ trend found outside the central region strengthens
our earlier conclusion (Section~\ref{sssec:TRdisc}) that
ram-pressure stripping, ram-pressure triggering of star formation,
tidal truncation and tidal-triggering of star formation by the
main cluster potential are not dominant. Further a T-$\Sigma$
relation is also found locally in non-concentrated systems, where
presumably the effects of the overall cluster potential and the
ICM are negligible; Dressler 1980).

Finally, Figure~\ref{fig:TScfr0} compares our T-$\Sigma$ relation
for Cl0024+16 with that determined locally (Dressler 1980, revised
in Dressler et al.\ 1997). At projected densities higher than
$\sim 20$ gal Mpc$^{-2}$, the fraction of early-type galaxies is
higher in the local Universe than in Cl0024+16 at any given
$\Sigma$. This agrees with other intermediate redshift studies
which indicate that the fraction of E+S0 declines with redshift
(e.~g. van Dokkum et al.\ 2001). An equally interesting result
(unique to the Cl0024+16 dataset) is that at local densities below
$\sim 20$ gal Mpc$^{-2}$ the morphological mix at $z\approx 0.4$
is indistinguishable (within the errors) to that observed locally.
A related result is seen in the T-R relation at the periphery
(c.f. Figure~\ref{fig:earlyfrac} with Figure 12 in Dressler et
al.\ 1997). The field number densities are insufficiently precise
to establish whether this is a displacement of E+S0 galaxies from
the field to the highest density regions, a transformation of
spirals into E+S0s, or a combination. Regardless, it is clear that
the morphological mix of galaxies in low density environments has
not changed significantly since $z\sim0.4$ ($\sim30$ \% of the
Hubble time). In combination with the (at most) mild gradients of
morphological fraction at large radii discussed in
Section~\ref{sssec:TRdisc}, and with the mild decline in star
formation rate in the outer parts of clusters, we interpret this
as further evidence that slow environmental processes are at work
in the outer parts of clusters.

\begin{inlinefigure}
\begin{center}
\resizebox{\textwidth}{!}{\includegraphics{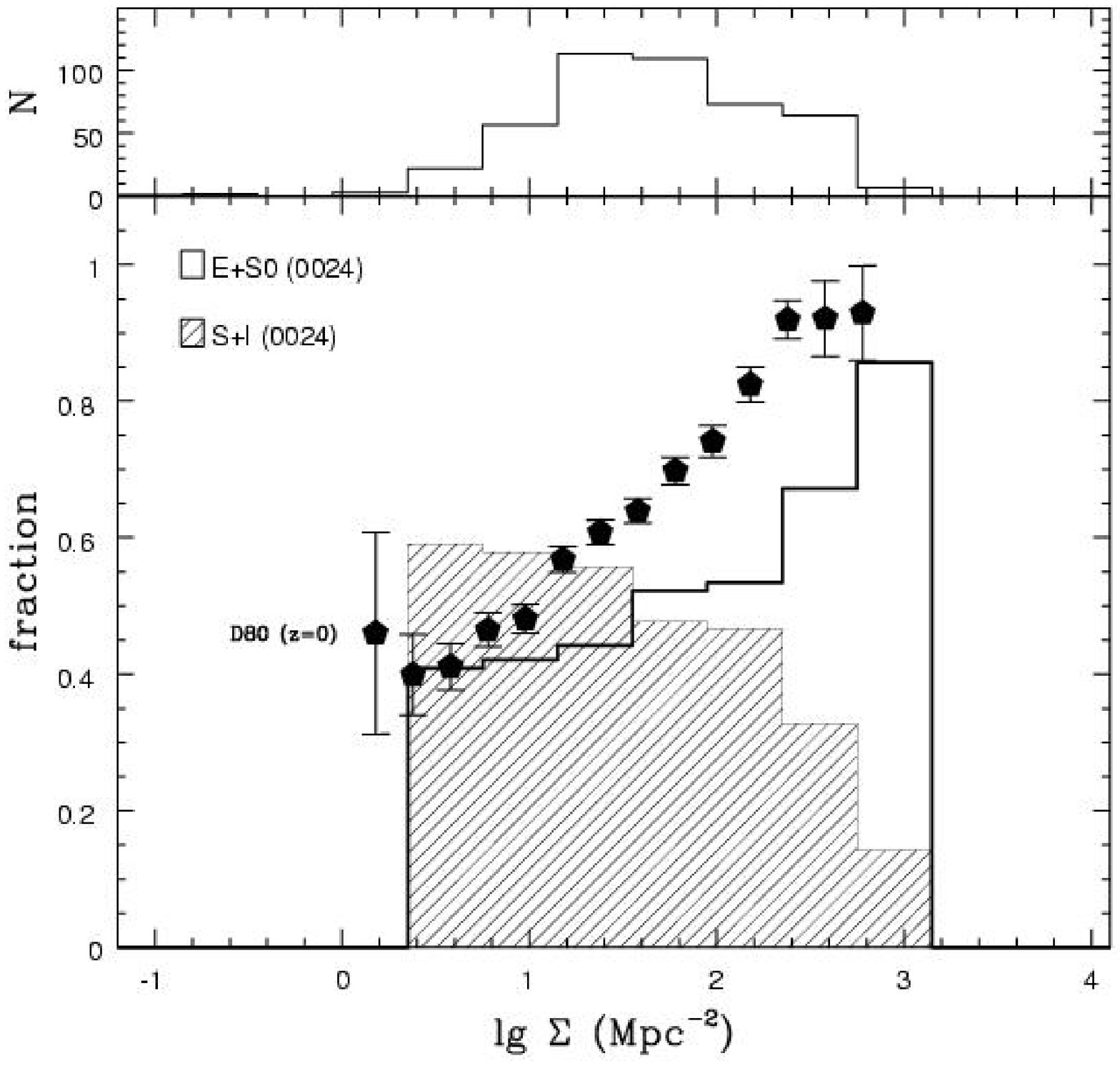}}
\end{center}
\figcaption{The morphology-projected density relation for galaxies
in Cl0024+16 (histogram) compared to the local relation as
determined by Dressler et al.\ (1997). The local early-type
fraction is shown as red solid pentagons (for clarity the spiral
fraction is omitted). The local relation has been corrected to our
adopted value of the Hubble constant $H_0=65$ \kms Mpc$^{-1}$.
\label{fig:TScfr0}}
\end{inlinefigure}

\section{Conclusions}

\label{sec:sum}

In this paper we have presented the first results from a wide
field HST survey of the rich galaxy cluster Cl0024+16 at
$z\approx0.4$. We exploit an unique wide-field mosaic of HST WFPC2
images and relate these to available spectroscopic data in order
to examine the morphological mix as a function of radius and
projected galaxian density. Our main observational results can be
summarized as follows:

\begin{enumerate}
\item We have obtained WFPC2 observations (39 independent pointings)
covering an approximately circular region of 10 Mpc in diameter around
the center of Cl0024+16. The areal sampling ranges from 100 \% at the
center to $\sim$20-40 \% at the periphery. We produce an object
catalog comprising over 22,000 object to $I\sim25$ (F814W). All $2200$
objects to $I=22.5$ have been morphologically classified in a scheme
broadly similar to that adopted for the Medium Deep Survey.

\item Approximately 100 new redshifts in the field have been obtained
using the W.~M. Keck I telescope. Together with other redshift
catalogs (Dressler et al.\ 1999; Czoske et al.\ 2001; Owen 2001;
Metevier \& Koo 2001) this brings the total number of objects with
redshift to 787 (including 358 members; $0.374<z<0.402$). After cross
correlation with the morphological WFPC2 catalog we created the
redshift WFPC2-z catalog, consisting of 362 objects with both redshift
and WFPC2 imaging (195 cluster members).

\item Using a simple model for the cluster based on the observed mass
estimate and X-ray surface brightness profile (Appendix A), we
have estimated the time and velocity scales for galaxy infall and
the regions of influence of various mechanisms proposed to account
for environment evolution. We define three projected regions; the
cluster center ($\la1$ Mpc) where tidal effects of the cluster
potential and those of the ICM are dominant; the transition region
($\sim$1-2.4 Mpc, comprising the virial radius) where galaxies do
not experience the strong interaction with the cluster potential
and the ICM, but some fraction of galaxies may have experienced
interaction with the cluster center in the recent past ($<1$ Gyr),
and the periphery beyond 2.4 Mpc where galaxies are entering the
cluster for the first time and where environmental trends are
driven by slow ($>1$ Gyr) phenomena.

\item We have used the morphological catalog (WFPC2) and the redshift
catalog (WFPC2-z) to study the galaxy population and the mix of
morphological types with cluster radius (the so-called $T-R$ relation)
to the previously unexplored regions beyond 1 Mpc at $z$=0.4. Cluster
galaxies ($0.374<z<0.402$) are found out to 5 Mpc from the cluster
center, but the marginal excess with respect to the background makes
precise studies difficult until further spectroscopy is available. The
fraction of members declines from $>$60-70\% in the center, to $\sim
40\%$ in the transition region, and less than $\sim20 \%$ at the
periphery. Consistent with previous investigations, the fraction of
E+S0 galaxies ($I<21.1$) is highest ($73\pm10$\%) in the central core
within 200 kpc from the center and declines rapidly to $\sim 50$ \% at
1 Mpc. It then changes very mildly over the next 4 Mpc reaching a
value ($43\pm 13$ \%) at the periphery indistinguishable from the
field ($37\pm7$ \%). The most significant changes in the $T-R$
relation occur within a radius of 1 Mpc.

\item We also studied the run of morphological mix with local
projected density (the so-called $T-\Sigma$ relation). At
intermediate redshift we extend this relation by over an order of
magnitude in $\Sigma$ over previous determinations (Dressler et
al.\ 1997). The fraction of E+S0 galaxies falls rapidly over the
densest decade and then flattens out at $\Sigma \sim 10$ gal
Mpc$^{-2}$ at $\sim 45$\%, slightly above the field value. Outside
the central $\sim 0.5$ Mpc we find a large scatter in values of
$\Sigma$ at given radius, with overdensity peaks associated with
identifiable clumps in the two-dimensional galaxy distribution.
Overdensities at large cluster radii have the same mix of
morphological types as the similar ones at small cluster radii.
Remarkably, in the region of overlapping densities the T$-\Sigma$
relation of these inner and outer regions are identical. In
comparison to local samples, CL0024 has a smaller fraction of E+S0
galaxies at densities above $\sim 20$ gal Mpc$^{-2}$ (in agreement
with previous studies), while at densities below this limit the
fraction of E+S0 is unchanged within the uncertainties.

\end{enumerate}

Although we are continuing to gather spectroscopic data, we
interpret our current results as follows. The mild gradients found
at large radii in the morphological mix cannot be explained by
mechanisms such as tidal triggering of star formation and tidal
stripping of the dark matter halos by the main cluster potential,
nor by ram pressure triggering of star formation, or ram pressure
stripping, which are ineffective in these regions. Instead we
suggest that other mechanisms such as starvation or harassment
operate over timescales of several Gyrs. Starvation seems a
promising candidate to explain the mild trends in the
morphological mix, since it would explain also the mild gradients
in the star formation rate seen by Abraham et al.\ (1996b; see
also Balogh et al.\ 1999 and Balogh, Navarro \& Morris 2000).

A gradual mechanism would also explain the homogeneity of the
early-type galaxy population located in the cluster centers
(Bower, Lucey \& Ellis 1992; Ellis et al. 1997; Stanford,
Eisenhardt \& Dickinson 1998; van Dokkum et al.\ 1998b; Kelson et
al.\ 2000), and the small differences observed between their
clustered and field populations at various redshifts (Bernardi et
al.\ 1998; Treu et al.\ 1999, 2001b, 2002; Kochanek et al.\ 2000;
van Dokkum et al.\ 2001). The homogeneity might be understood if
the quenching of star formation occurred gradually rather than via
a series of discrete events. Secondary bursts of star formation
can be reconciled with the homogeneity of the stellar populations
in local and distant E+S0 galaxies (e.g. Bower, Kodama \&
Terlevich 1998; Treu et al.\ 2001b) and with the detection of
post-starburst K+A galaxies in clusters (Poggianti et al.\ 1999).
However, if the mass distribution was also altered -- as during
harassment -- significant fine-tuning of the star formation and
structural changes would be needed to preserve the tight scaling
laws. High quality spectroscopic data, secured as a function of
cluster radius, will address these issues and help to identify the
mechanisms at work. Such data will also help clarify the origin of
the sharp rise in the E+S0 fraction in the core.

A more attractive (not necessarily alternative) possibility is
that the T-R relation is mostly due to segregation, following an
inbuilt correlation between galaxy properties and their local
substructures. This is supported by a convincing T-$\Sigma$
relation outside the central region where $\Sigma$ and radius show
a large scatter. In order for the T-$\Sigma$ relation to be
sustained outside the central concentration, not only must such
correlations be defined within each subclump, but also the
galaxies contained within them must be moving coherently. Although
a more detailed physical picture must await further dynamical data
in the cluster outskirts, we present an emerging picture where the
assembly and star formation of individual subclumps falling into
the cluster (see also Kodama et al.\ 2001) dominates the
morphological trends outwards of $\sim 0.5-1$ Mpc.

\acknowledgments

We are grateful to Frazier Owen, Anne Metevier, and David Koo for
sharing their lists of redshifts with us in advance of publication.
We thank Judy Cohen and Patrick Shopbell for writing and supporting
the software {\sc autoslit} that we used to design the LRIS masks. TT
acknowledges useful conversations with Andrew Benson, Kevin Bundy,
Chris Conselice, Bianca Poggianti, David Sand, Pranjal Trivedi, Pieter
van Dokkum. The referee is thanked for comments that helped clarifying
the presentation of the results.  Finally, the authors wish to
recognize and acknowledge the cultural role and reverence that the
summit of Mauna Kea has always had within the indigenous Hawaiian
community.  We are most fortunate to have the opportunity to conduct
observations from this mountain.

This paper is based on observations collected at the W.~M. Keck
Observatory, which is operated jointly by the California Institute of
Technology and the University of California, and with the NASA/ESA
Hubble Space Telescope, obtained at the Space Telescope Science
Institute, which is operated by AURA, under NASA contract NAS5-26555.
We acknowledge financial support for proposal number HST-GO-8559
provided by NASA through a grant from STScI, which is operated by
AURA, under NASA contract NAS5-26555. Ian Smail acknowledges support
from the Royal Society and the Leverhulme Trust.

\appendix

\section{A simple cluster model}

\label{app:model}

In this appendix we consider a very simple model for a galaxy
infalling radially onto the cluster. This model will be later used to
estimate the relevant spheres of influence of physical mechanisms
operating in Cl0024+16 and the infalling timescales.

First, we need to compute the {\it virial radius} $r_{\rm V}$ which
contains the virialized mass (Gunn \& Gott 1972). This can be obtained
in terms of the galaxy velocity dispersion at large radii
($\sigma_{\infty})$ as
\begin{equation}
r_{\rm V} =
        1.7\,h^{-1}\,\mathrm{Mpc}\,
    \left(\frac{\sigma_\infty}{1000\,\mathrm{km\,s^{-1}}}\right)\,
        \left[\Omega_{\rm m}(1+z)^3 + \Omega_\Lambda\right]^{-1/2}
\end{equation}

\noindent in a flat cosmology (Carlberg, Yee, Ellingson 1997;
Girardi \& Mezzetti 2001), i.e. $\sim 2$ Mpc adopting
$\sigma_{\infty}=911$ \kms (Girardi \& Mezzetti 2001) with our
choice of the cosmological parameters. For Cl0024+16, the velocity
dispersion is affected by the presence of a secondary peak in the
distribution suggestive of an infalling group (Czoske et al.\
2002). At large radii the velocity dispersion of primary peak
flattens at $\sim 600$ \kms, and this can be considered a lower
limit to $\sigma_{\infty}$, hence $r_{\rm V}\ga1.3$ Mpc. For the
present analysis, we will consider the indicative value of the
virial radius to be approximately $r_{\rm V}\sim1.7$ Mpc.

For simplicity, we assume the cluster mass density scales as
$\rho\propto r^{-2}$ inside the $r_{\rm V}$ and is zero outside. The
absolute normalization is obtained using the values given by Soucail
et al.\ 2001 (we consider the average of X-ray and strong lensing
normalization, with a factor of two uncertainty).  A radially
infalling galaxy starting with zero velocity at turnaround radius
$r_a$ hits the cluster at $r_{\rm V}$ with velocity $V_{\rm V}$, as
given by energy conservation
\begin{equation}
V_{\rm V}=\sqrt{\frac{2GM_{\rm V}}{r_{\rm V}}-\frac{2GM_{\rm V}}{r_{\rm a}}},
\label{eq:Energy}
\end{equation}

where G is the gravitational constant and $M_{\rm V}$ is the virial
mass.  For simplicity we take $r_a/r_{\rm V}=2$ (the uncertainty is
dominated by other factors, such as mass uncertainty, using 1.5,3 or 4
would not alter significantly the picture; for a more realistic
description of the orbital properties of the infalling population see,
e.g., Vitvitska et al.\ 2002), obtaining
\begin{equation}
V_{\rm V}=\sqrt{\frac{GM_{\rm V}}{r_{\rm V}}}.
\label{eq:Energy2}
\end{equation}

Using $M_{\rm V}=8\cdot 10^{14} M_{\odot}$ and $r_{\rm V}=1.7$ Mpc, we obtain
$V_{\rm V}=1400 $\kms. 

\begin{inlinefigure}
\begin{center}
\resizebox{\textwidth}{!}{\includegraphics{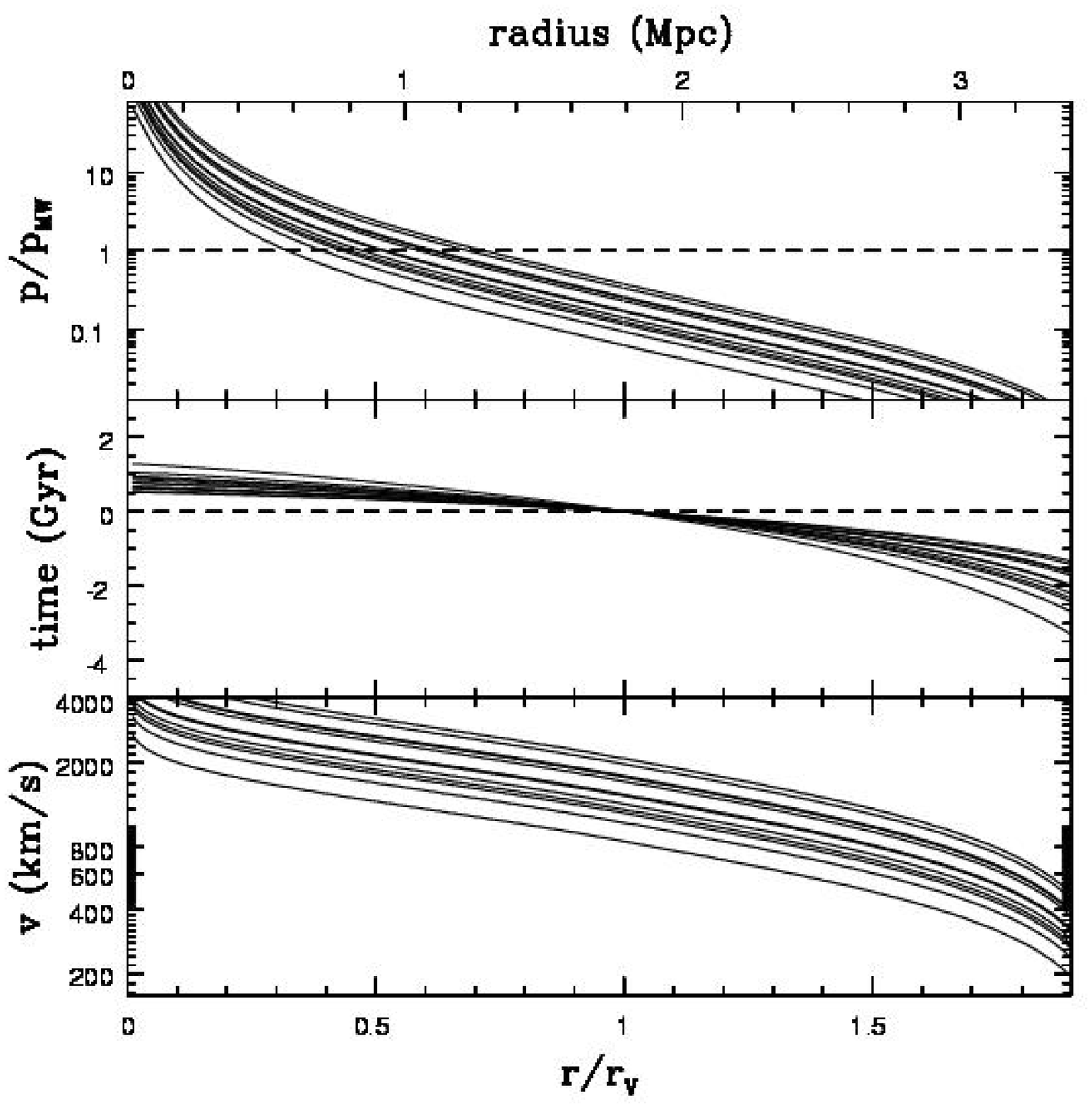}}
\end{center}
\figcaption{A radially infalling galaxy. Lower panel: velocity of a
radially infalling galaxy computed using a static isothermal model for
the mass distribution of cluster CL0024+16. Middle panel: travel time
from the virial radius. Upper panel, ram pressure in units of the ram
pressure needed to strip a galaxy like the Milky-Way of its gas. The
various lines are otained for different values of the cluster mass
normalization (Soucail et al.\ 2000), and of the ratio between
turnaround radius and virial radius $r_{ta}/r_{\rm V}=1.5,2,3,4$.
\label{fig:ram}}
\end{inlinefigure}

The infalling velocity $v_{\rm i}(r)$ and time $t_{\rm i}(r)$ as a
function of radius are then found solving the equation of motion,
providing a rough estimate of the motion of infalling galaxies (see
Fig~\ref{fig:ram}. The travel time between the center and the virial
radius is $\sim 1$ Gyr, while 2-4 Gyr are needed to travel between the
$r_{\rm V}$ and 2 $r_{\rm V}$ (depending on mass normalization and on
$r_{\rm a}/r_{\rm V}$). These travel times imply average velocities of
order 2000 \kms\, between the center and $r_{\rm V}$ and of order
$500-1000$ \kms\, between $r_{\rm V}$ and $2r_{\rm V}$. In general, we
expect a continuous distribution of orbital parameters for the
infalling galaxies with only a small fraction of galaxies to be on
highly eccentric or quasi-radial orbits (for example Ghigna et al.\
1998 found that $\sim 20$ \% of the galaxies will pass in the central
200 kpc, i.e. $\sim r_{\rm V}/10$). Hence we can assume that the
travel times estimated with radial infalling galaxies are
approximately a lower limit to the travel times of the population.

\section{Length scales}

\label{app:ls}

The first important quantity is the {\it stripping radius} $r_{\rm
st}$ defined as that where a radially infalling galaxy with the
properties of the Milky Way would have its gas removed by
ram-pressure. We can estimate the stripping radius by considering the
cluster mass model introduced in~\ref{app:model} together with the
properties of the inter cluster medium.  The gas density profile
inside $r_{\rm V}$ is modeled with a $\beta$ profile
\begin{equation}
\rho_{gas}=\rho_{0}\left[1+\left(\frac{r}{r_c}\right)^2\right]^{-\frac{3}{2}\beta},
\label{eq:beta}
\end{equation}

\noindent
with $\beta=0.475$, $r_c=60$ kpc and
$\rho_{0}=3\cdot10^{14}M_{\odot}$Mpc$^{-3}$ (B\"ohringer et al.\
2001).  Assuming the X-ray emitting gas is at rest with respect to the
cluster, we can directly compute the ram pressure $p=\rho_{gas}v_i^2$
and check where ram pressure stripping is effective using the
condition (Gunn \& Gott 1972; see also Abadi, Moore \& Bower 1999,
Fujita \& Nagashima 1999; Fujita 2001; Toniazzo \& Schindler 2001)
\begin{equation}
\rho_{gas}v_i^2>2.1\cdot 10^{-12} {\rm N m}^{-2} \left(\frac{v_{rot}}{220 {\rm km
s}^{-1}}\right)^2\left(\frac{r_h}{10
{\rm kpc}}\right)^{-1}\left(\frac{\Sigma_{HI}}{8\cdot10^{20} m_H
{\rm cm}^{-2}}\right),
\label{eq:ram}
\end{equation}
where the rotational velocity of the infalling galaxy $v_{rot}$, its
scale length ($r_h$), and surface density of HI ($\Sigma_{HI}$), are
expressed in units appropriate for the Milky-Way (Spitzer 1978; Binney
\& Tremaine 1987). For an infalling Milky-Way the stripping radius
occurs in the range $r_{st}=0.5-1$ Mpc, therefore we will assume
$r_{st}\sim0.75$Mpc. A similar estimate is obtaining by considering
that -- given central velocity dispersion of the cluster galaxies
$\sim 1000$ \kms\, (Soucail et al.\ 2001) -- galaxies in the center of
the cluster will have maximal velocity with respect to the X-ray
emitting gas of order $\sim 2000$ \kms. Using the gas density given in
Equation~\ref{eq:beta} and the condition in Equation~\ref{eq:ram}, we
obtain a stripping radius $\sim 0.8$ Mpc. Note that we are considering
here the most favorable conditions for ram pressure stripping
(i.e. maximal velocity) so that $r_{st}$ is practically the maximal
radius at which we expect to ram-pressure stripping to be effective.
In summary, we estimate that for Cl0024+16 $r_{\rm st}=0.5-1$ Mpc at
most.

Alternate mechanisms (thermal evaporation, viscous and turbulent
stripping) might be more effective in this respect, depending on the
detailed physics and the galactic orbits. For simplicity, we will
assume that alternate mechanisms occur at most with the same intensity
and therefore do not affect our estimate of the region where stripping
is important.

The extended gas reservoir that may surround field galaxies (see
Benson et al.\ 2000), sustains star formation in low density
environments and is more loosely bound. Accordingly it can be
removed at radii beyond $r_{st}$. This effect is dependent on the
detailed physics, the galaxy orientation and the structure of the
diffuse reservoir. We will assume that such starvation can be also
effective at least to the virial radius (e.~g. Abadi et al.\ 1999;
Balogh et al.\ 2000).

Interactions with the ICM can also trigger star formation by
increasing pressure on the ISM. However, according to Fujita (1998)
and Fujita \& Nagashima (1999), this mechanism seems to be negligible
compared with tidally-induced pressure or harassment-induced pressure
that we describe below.

Tidal interactions with the cluster potential affect galaxies in
several ways. The effectiveness of tidal compression in triggering
star formation depends on the cluster mass distribution, but
detailed calculations (Byrd \& Valtonen 1990; Herinksen \& Byrd
1996; Fuijta 1998) show that for clusters of mass comparable to
Cl0024+16 the process is only effective within the central $\sim
200$ kpc at most. Tidal stripping (Merritt 1983, 1984) removes
mass from the outer regions of a galaxy (typically at radii larger
than the luminous component), thus altering the mass and cross
sections for galaxy-galaxy interactions. Modeling the cluster and
galaxy halos as isothermal spheres with velocity dispersions
$\sigma_{\rm c}$ and $\sigma_{\rm h}$ (Moore et al.\ 1996, 1998)
we derive the following estimate for the {\it cluster tidal
truncation radius} $r_{\rm tt}$
\begin{equation}
r_{\rm tt} \ge \frac{\sigma_c}{\sigma_h} \sim r_{\rm o}
\left( \frac{\sigma_{\rm h}}{1000\, \mathrm{km\,s^{-1}}}\right)^{-1},
\label{eq:tidal3}
\end{equation}
limiting the cluster region where tidal truncation is important for
the luminous component of the galaxies inside their optical radius
$r_{\rm o}$.

Less dense galaxies will be disrupted first, but typically only in
the innermost regions (100-200 kpc; Moore et al.\ 1998). Removal
of the diffuse gas reservoir will also be dominated by
hydrodynamical effects and therefore we will neglect this
mechanism in the estimation of regions where starvation is
important.

Finally, we consider galaxy-galaxy interactions. Harassment (Moore et
al.\ 1996) arises from an interplay between the overall cluster
potential -- which is responsible for tidal truncation of halos and
the high speed of galaxies -- and the local density -- which affects
the interaction rate. Schematically, the harassment rate $f_H$ scales
with the luminous galaxy density $\rho_{gal}$ and mass m$_{gal}$ as
\begin{equation}
f_H\propto \rho_{gal} m^2_{gal}\propto \rho_{gal} r^{2},
\label{eq:harass}
\end{equation}

\noindent where we assume that galaxy masses are tidally truncated
by the cluster potential. In the simple case where local density
scales smoothly as $r^{-2}$ -- the harassment rate should be
independent on cluster radius, within the validity limits of
Equation~\ref{eq:harass} (Moore et al.\ 1998). In practice, there
is spread in local density at a given radius (see Section~6), and
therefore the rate depends somewhat on both variables. Note the
effects of harassment on the morphology of a galaxy depend not
only on its mass, but also on its internal structure. For example,
typical $L_*$ spirals in Cl0024+16 could be transformed either
into dwarf spheroidals or into lenticulars depending on the
concentration of the mass distribution and on the disk scale
length (Moore et al.\ 1999).

The merger frequency is a strong function of cluster radius, since
it increases with density but decreases with velocity dispersion.
Schematically, the rate peaks around the virial radius of the
cluster and declines towards the center (see e.g. Ghigna et al.\
1998). In hierarchical clustering, galaxies located at any given
time at the cluster center are those that formed earlier and had
the highest probability of having undergone major mergers in their
past (Springel et al. 2001; Diaferio et al. 2001). This
distinction emphasizes the difference between regions where
phenomena happen from the regions where their effects are
observed, as discussed in Section~\ref{ssec:2D}.

The regions of the cluster where the above physical processes are most
effective are summarized in the upper panel of
Figure~\ref{fig:mechanisms}.

\section{Temporal and Velocity Scales}

\label{app:tvs}

To interpret the observed distribution of galaxies across a wide
range of cluster radius we also need to consider the time scales
involved and thus, by implication, the galaxy motions.

First, we estimate the time scales associated with the motions of
galaxies in the cluster potential. As described in
Appendix~\ref{app:model}, for a galaxy on
a radial orbit, approximately\footnote{Note that this simple model
assumes a fixed cluster potential, so it is clearly not appropriate
for timescales of several Gyrs, where we expect the cluster to evolve
significantly, and therefore should only be regarded as indicative.},
$\sim 1$ Gyr is needed to traverse from the center to the virial
radius, and $\sim 2-4$ Gyrs to travel between $r_{\rm V}$ and $2
r_{\rm V}$. As galaxies are not on purely radial orbits (e.~g. Ghigna
et al.\ 1998; van der Marel et al.\ 2000; Vitviska et al.\ 2002), a
significant fraction of galaxies at or beyond the virial radius will
never reach the cluster center.  Hence these travel times can be
considered lower limits. For motions in any given direction, the
relevant velocity scale is given by $\sqrt{2GM_{\rm V}/r_{\rm V}}$
(where $M_{\rm V}$ is the virial mass, see Appendix~\ref{app:model}),
corresponding to velocities of order $\sim 1$ Mpc Gyr$^{-1}$.

Second, a time scale can be associated with each of the {\em physical
mechanisms} listed in Section~\ref{ssec:zoo}. Starting with
interactions between galaxies and the ICM, ram pressure stripping in
the highest density ICM regions happens on a very short time scale
($\sim 5\times 10^7$ yr; Abadi et al.\ 1999), while starvation
according to our definition occurs more slowly, lasting up to several
Gyrs. Galaxy-cluster gravitational interactions and tidal compression
of gas clouds occur rapidly ($\sim 10^8$ yr), while tidal truncation
occurs on time scales longer than the cluster crossing time, i.e. a
few Gyr . Harassment transforms the morphology and star formation
properties on timescales of a few Gyrs (e.g. Moore et al.\ 1999) while
mergers can take place on time scales of a fraction of a Gyr and their
morphological remnants would be undetectable after a few $10^8$ yr
(e.~g. Mihos 1995).

Third, we can consider the timescales for witnessing changes in the
star formation properties of a system. After a burst of star
formation, massive stars are dominant at most during the initial $\sim
10^8$ yr while signatures such as strong Balmer emission lines such be
detectable for $\sim 1$ Gyr after the burst (as k+a/a+k if the burst
happened on top of a older stellar population; Dressler \& Gunn 1983;
Poggianti et al.\ 1999). The exact value depends not only on the prior
and current star formation details but also on the actual observable
(e.g. Barger et al.\ 1996). However, for simplicity in ths paper we
will adopt a typical time of $\sim 1$ Gyr for alteration in the colors
and luminosity of a galaxy after a burst.

Finally, the timescale associated with starvation is linked to
that for gas consumption due to star formation. Larson et al.\
(1980) estimate that it takes a few Gyrs for a typical Milky Way
Galaxy to run out of gas, and cease star formation. Thereafter,
they show that the optical colors change slowly after the next few
Gyrs (see also Abraham et al.\ 1996b).

\begin{deluxetable}{lcc}
\tabletypesize{\scriptsize}
\tablecaption{Summary of relevant observed
quantities for cluster CL0024+16. \label{tab:CL0024+16}}
\tablewidth{0pt}
\tablehead{
\colhead{Observable} & \colhead{Value} & \colhead{Ref}}

\startdata
r$_c\tablenotemark{a}$ (X-ray)             & $10\farcs4$                                 & (1) \\
$\beta\tablenotemark{b}$ (X-ray)           & $0.475$                                     & (1) \\
L$_X$ (X-ray)                              & $3.7 \cdot 10^{44}$ erg s$^{-1}$            & (2) \\
T$_X$ (X-ray)                              & $5.7^{+4.9}_{-2.1}$ keV                     & (2) \\
center (X-ray)                             & 00:26:36.3 17:09:46                         & (2) \\
center (galaxies)                          & 00:26:34.8 17:10:05                         & (4) \\
center (BCG)                               & 00:26:35.7 17:09:43                         & (0) \\
$\sigma_{\infty}$                          & 911 \kms                                    & (4) \\
R$_{vir}$                                  & 1.7 Mpc                                     & (0) \\
Concentration                              & 0.53                                        & (3) \\
\enddata
\tablenotetext{a}{Core radius of the best-fitting $\beta$-model.}

\tablecomments{All coordinates are J2000. Data from (0) this paper ;
(1) B\"ohringer et al.\ (2000); (2) Soucail et al.\ (2000); (3)
Dressler et al.\ 1997; (4) Girardi \& Mezzetti 2001}

\end{deluxetable}

\begin{deluxetable}{cccccccc}
\tabletypesize{\scriptsize}
\tablecaption{Summary of pointings \label{tab:pointings}}
\tablewidth{0pt}
\tablehead{
\colhead{POS} & \colhead{RA(J2000)}   & \colhead{DEC(J2000)}   &
\colhead{PA\_V3} &
\colhead{exptime$\tablenotemark{a}$}  & \colhead{Ngal$\tablenotemark{b}$} & \colhead{Nmorph$\tablenotemark{c}$} &
\colhead{Nz}}
\startdata
00 & 00:26:37.4 & +17:08:54.7 & 148.29 & 19800 & 508 & 188 & 77   \\
01 & 00:27:01.5 & +17:18:49.8 & 280.00 & 4400 &  269 & 41  & 2    \\
02 & 00:26:47.0 & +17:19:42.3 & 280.00 & 4400 &  266 & 51  & 5    \\
03 & 00:26:32.6 & +17:20:34.6 & 280.00 & 4400 &  262 & 44  & 4    \\
04 & 00:27:12.3 & +17:14:30.2 & 280.00 & 4400 &  267 & 31  & 3    \\
05 & 00:26:56.8 & +17:15:24.3 & 61.84  & 4000 &  279 & 55  & 8    \\
06 & 00:26:46.4 & +17:16:12.3 & 239.09 & 4400 &  280 & 58  & 17   \\
07 & 00:26:27.8 & +17:17:09.0 & 62.04  & 4000 &  252 & 61  & 16   \\
08 & 00:26:13.4 & +17:18:01.3 & 62.67  & 4000 &  249 & 53  & 6    \\
09 & 00:27:23.2 & +17:10:15.5 & 244.20 & 4400 &  256 & 46  & 0    \\
10 & 00:27:08.4 & +17:10:59.2 & 314.99 & 4400 &  286 & 45  & 3    \\
11 & 00:26:53.1 & +17:11:57.1 & 62.73  & 4000 &  270 & 54  & 11   \\
12 & 00:26:39.8 & +17:12:52.4 & 246.80 & 4400 &  275 & 51  & 17   \\
13 & 00:26:24.2 & +17:13:41.5 & 61.36  & 4000 &  292 & 82  & 17   \\
14 & 00:26:10.8 & +17:14:32.0 & 280.00 & 4400 &  296 & 49  & 6    \\
15 & 00:25:56.4 & +17:15:29.5 & 243.13 & 4400 &  395 & 53  & 0    \\
16 & 00:27:18.9 & +17:06:46.7 & 241.11 & 4400 &  331 & 61  & 0    \\
17 & 00:27:04.5 & +17:07:39.1 & 240.87 & 4400 &  350 & 60  & 0    \\
18 & 00:26:53.2 & +17:08:46.9 & 240.31 & 4400 &  290 & 48  & 13   \\
19 & 00:26:21.3 & +17:10:08.2 & 323.79 & 4400 &  280 & 58  & 15   \\
20 & 00:26:06.1 & +17:11:08.0 & 73.00  & 4000 &  226 & 36  & 4    \\
21 & 00:25:51.6 & +17:11:58.8 & 63.42  & 4000 &  299 & 54  & 0    \\
22 & 00:27:14.8 & +17:03:14.1 & 35.68  & 4400 &  319 & 54  & 3    \\
23 & 00:26:45.8 & +17:05:02.5 & 63.18  & 4000 &  292 & 49  & 3    \\
24 & 00:26:32.5 & +17:05:52.9 & 280.00 & 4400 &  288 & 48  & 9    \\
25 & 00:26:16.9 & +17:06:46.9 & 62.32  & 4000 &  263 & 48  & 10   \\
26 & 00:26:02.5 & +17:07:39.1 & 62.32  & 4000 &  273 & 42  & 4    \\
27 & 00:25:48.2 & +17:08:29.0 & 39.49  & 4000 &  285 & 51  & 0    \\
28 & 00:26:56.6 & +17:00:42.5 & 61.29  & 4000 &  273 & 62  & 5    \\
29 & 00:26:27.7 & +17:02:27.2 & 61.45  & 4000 &  293 & 40  & 12   \\
30 & 00:26:14.4 & +17:03:17.7 & 280.00 & 4400 &  309 & 63  & 10   \\
31 & 00:25:58.8 & +17:04:11.8 & 62.96  & 4000 &  290 & 50  & 4    \\
32 & 00:26:41.1 & +16:58:37.1 & 241.36 & 4400 &  341 & 76  & 3    \\
33 & 00:26:24.7 & +16:59:01.8 & 240.07 & 4400 &  270 & 41  & 3    \\
34 & 00:26:10.2 & +16:59:54.0 & 241.06 & 4400 &  325 & 55  & 8    \\
35 & 00:26:03.3 & +17:18:53.0 & 62.79  & 4000 &  301 & 69  & 6    \\
36 & 00:26:30.8 & +17:11:28.9 & 295.39 & 4400 &  357 & 108 & 40   \\
37 & 00:26:43.6 & +17:07:44.8 & 248.60 & 4400 &  342 & 61  & 12   \\
38 & 00:26:29.9 & +17:08:44.6 & 62.95  & 4000 &  384 & 111 & 40   \\
\enddata

\tablenotetext{a}{Total exposure time in seconds}
\tablenotetext{b}{To $I=25$}
\tablenotetext{c}{To $I=22.5$}

\tablecomments{For each pointing we list position, orientation,
exposure time, number of objects detected, number of objects with
morphological classification, and number of spectroscopic redshifts}

\end{deluxetable}

\begin{deluxetable}{ll}
\tabletypesize{\scriptsize}
\tablecaption{Summary of entries in the catalog \label{tab:entries}}
\tablewidth{0pt}
\tablehead{
\colhead{Entry} & \colhead{Description}}
\startdata
POS &   WFPC2 pointing, as identified in Table~2 \\
chip    &   WFPC2 chip \\
ID  &   running number within each chip \\
X   &   x-coordinate on the chip of the object baricenter\\
Y   &   y-coordinate on the chip of the object baricenter\\
RA  &   Right Ascension (J2000) of the object baricenter\\
DEC &   Declination (J2000) of the object baricenter\\
IAU &   F814W mag\_auto (SExtractor) \\
$\delta$IAU & error on IAU (SExtractor) \\
IAP1    &   F814W magnitude within $0\farcs5$ diameter aperture (SExtractor) \\
$\delta$IAP1    & error on IAP1 (SExtractor) \\
IAP2    &   F814W magnitude within $1''$ diameter aperture (SExtractor) \\
$\delta$IAP2    & error on IAP2 (SExtractor) \\
star    &   class\_star (SExtractor) \\
T   &   morphological type (RSE) \\
Comment &   Comment by RSE\\
\enddata
\end{deluxetable}

\tablecomments{Standard SExtractor (Bertin \& Arnouts 1996) parameters
are indicated by their standard name, and a full description can be
found in SExtractor user's handbook.}

\begin{deluxetable}{ccrrrrrrrrcrl}
\tabletypesize{\tiny}
\tablecaption{First entries of the catalog \label{tab:catalog}}
\tablewidth{0pt}
\tablehead{
\colhead{POS} & \colhead{chip} & \colhead{ID} & \colhead{X} & \colhead{Y} & \colhead{RA} & \colhead{DEC} & \colhead{IAU$\pm$ $\delta$ IAU} & \colhead{IAP1$\pm$ $\delta$ IAP1} & \colhead{IAP2 $\pm$ $\delta$ IAP2} & \colhead{star} & \colhead{T} & \colhead{
Comment}   }
\startdata
0 & 2 & 1 & 1370.99 & 57.35    &   6.65112 &  17.131149 & 21.344$\pm$0.006  & 24.385$\pm$0.016 & 23.528$\pm$0.013 & 0.0 & 9 & -    \\
0 & 2 & 2 & 1476.62 & 8.03     &   6.65148 &  17.129572 & 21.641$\pm$0.005  & 23.141$\pm$0.007 & 22.007$\pm$0.004 & 0.0 & 9 & -    \\
0 & 2 & 3 & 1079.62 & 28.92    &   6.65243 &  17.135000 & 23.953$\pm$0.030  & 24.890$\pm$0.023 & 24.240$\pm$0.023 & 0.0 & - & -    \\
0 & 2 & 4 & 114.58  & 21.22    &   6.65556 &  17.148020 & 23.801$\pm$0.025  & 25.039$\pm$0.026 & 24.112$\pm$0.021 & 0.0 & - & -    \\
0 & 2 & 5 & 46.78   & 19.46    &   6.65580 &  17.148932 & 23.012$\pm$0.012  & 23.924$\pm$0.012 & 23.282$\pm$0.011 & 0.0 & - & -    \\
0 & 2 & 6 & 1441.84 & 16.16    &   6.65148 &  17.130072 & 24.393$\pm$0.045  & 26.099$\pm$0.062 & 25.119$\pm$0.049 & 0.0 & - & -    \\
0 & 2 & 7 & 205.43  & 3.96     &   6.65552 &  17.146742 & 24.424$\pm$0.032  & 25.028$\pm$0.025 & 24.597$\pm$0.027 & 0.0 & - & -    \\
0 & 2 & 8 & 1321.63 & 9.76     &   6.65195 &  17.131672 & 22.189$\pm$0.011  & 24.743$\pm$0.021 & 23.298$\pm$0.011 & 0.0 & 4 & edge \\
\enddata
\end{deluxetable}

\tablecomments{Head of the WFPC2 catalog of CL0024+16. The first lines
are available for guidance, the entire catalog of is available in
electronic format. See Table 3 and text for description of entries.}

\end{document}